\documentclass{IEEEcsmag}
\usepackage{upmath}
\usepackage{balance}
\usepackage[acronym,nonumberlist,nowarn]{glossaries}
    \glsdisablehyper
    \newacronym{ADI}{ADI}{Alternating Direction Implicit}
\newacronym{AGU}{AGU}{Address Generation Unit}
\newacronym[longplural={Access History Tables}]{AHT}{AHT}{Access History Table}
\newacronym{AHT-R}{AHT-R}{Access History Table - Read}
\newacronym{AHT-W}{AHT-W}{Access History Table - Write-Back}
\newacronym{AIP}{AIP}{Access Interval Predictor}
\newacronym{AL}{AL}{Added Latency for column accesses}
\newacronym{ASIC}{ASIC}{Application Specific Integrated Circuit}
\newacronym{AVX}{AVX}{Advanced Vector Extensions}
\newacronym[longplural={Basic Block Vectors}]{BBV}{BBV}{Basic Block Vector}
\newacronym{BHT}{BHT}{Branch History Table}
\newacronym{HBM}{HBM}{Hybrid Bandwidth Memory}
\newacronym{DDR4}{DDR4}{Double-Data Rate 4}

\newacronym{NN}{NN}{Neural network}
\newacronym{BTB}{BTB}{Branch Target Buffer}
\newacronym{CAS}{CAS}{Column Address Strobe}
\newacronym{AMAT}{AMAT}{Average Memory Access Time}
\newacronym{CCD}{CCD}{Column to Column Delay}
\newacronym{AI}{AI}{Arithmetic Intensity}
\newacronym[longplural={Columnar Database Systems}]{CDBS}{CDBS}{Columnar Database System}
\newacronym[longplural={Chip Multiprocessors}]{CMP}{CMP}{Chip Multiprocessor}
\newacronym{CMT}{CMT}{Chip Multithreading}
\newacronym[longplural={Coarse-Grain Reconfigurable Arrays}]{CGRA}{CGRA}{Coarse-Grain Reconfigurable Array}
\newacronym{C-RAM}{C-RAM}{Computational-RAM}
\newacronym{CWD}{CWD}{Column Write Delay}
\newacronym{DBI}{DBI}{Dynamic Binary Instrumentation}
\newacronym[longplural={Database Management Systems}]{DBMS}{DBMS}{Database Management System}
\newacronym{DDR}{DDR}{Double Data Rate}
\newacronym{DEWP}{DEWP}{Dead Line and Early Write-Back Predictor}
\newacronym{DRAM}{DRAM}{Dynamic Random Access Memory}
\newacronym{DSBP}{DSBP}{Dead Sub-Block Predictor}
\newacronym{DSM}{DSM}{Decomposed Storage Manage}
\newacronym{FAW}{FAW}{Four row Activation Window}
\newacronym{FP}{FP}{Floating-Point}
\newacronym{EDP}{EDP}{Energy-Delay Product}
\newacronym{FCFS}{FCFS}{First-Come First-Serve}
\newacronym{FIFO}{FIFO}{Fist In, First Out}
\newacronym{FSM}{FSM}{Finite State Machine}
\newacronym{FPGA}{FPGA}{Field-Programmable Gate Array}
\newacronym[longplural={Functional Units}]{FU}{FU}{Functional Unit}
\newacronym{GAg}{GAg}{Global Adaptive branch prediction using one Global PHT}
\newacronym{GAs}{GAs}{Global Adaptive branch prediction using per-Set PHT}
\newacronym{GCC}{GCC}{GNU Compiler Collection}
\newacronym{GEMS}{GEMS}{General Execution-driven Multiprocessor Simulator}
\newacronym[longplural={General Purpose Processors}]{GPP}{GPP}{General Purpose Processor}
\newacronym{HMC}{HMC}{Hybrid Memory Cube}
\newacronym{HIVE}{HIVE}{HMC Instruction Vector Extensions}
\newacronym{IATAC}{IATAC}{Inter-Access Time per Access Count}
\newacronym{ILP}{ILP}{Instruction Level Parallelism}
\newacronym{IPC}{IPC}{Instructions per Cycle}
\newacronym{ISA}{ISA}{Instruction Set Architecture}
\newacronym{IRAM}{IRAM}{Intelligent RAM}
\newacronym{KIPS}{KIPS}{Kilo Instructions per Second}
\newacronym{LDS}{LDS}{Linked Data Structure}
\newacronym{LFSR}{LFSR}{Linear Feedback Shift Register}
\newacronym{LFMR}{LFMR}{Last-to-First Miss-Ratio}
\newacronym{MLP}{MLP}{Memory-Level Parallelism}

\newacronym{LLC}{LLC}{last-level cache}
\newacronym{LRU}{LRU}{Least Recently Used}
\newacronym{LSU}{LSU}{Load Store Unit}
\newacronym{LTP}{LTP}{Last-Touch Predictor}
\newacronym{LvP}{LvP}{Live-time Predictor}
\newacronym{LWP}{LWP}{Last Write Predictor}
\newacronym{MARSS}{MARSS}{Micro Architectural and System Simulator}
\newacronym{McPAT}{McPAT}{Multi-core Power, Area, and Timing}
\newacronym{MIPS}{MIPS}{Microprocessor without Interlocked Pipeline Stages}
\newacronym{MOB}{MOB}{Memory Order Buffer}
\newacronym{MMU}{MMU}{Memory Management Unit}
\newacronym{MPKI}{MPKI}{Misses per Kilo-Instruction}
\newacronym{MSHR}{MSHR}{Miss-Status Handling Registers}
\newacronym{NAS}{NAS}{Numerical Aerodynamic Simulation}
\newacronym{NDP}{NDP}{Near-Data Processing}
\newacronym{NMP}{NMP}{Near Memory Processor}
\newacronym{NoC}{NoC}{Network-on-Chip}
\newacronym{NPB}{NPB}{NAS Parallel Benchmark}
\newacronym{NUCA}{NUCA}{Non-Uniform Cache Architecture}
\newacronym{NUMA}{NUMA}{Non-Uniform Memory Access}
\newacronym{OoO}{OoO}{Out-of-Order}
\newacronym{OpenMP}{OpenMP}{Open Multi-Processing}
\newacronym{OS}{OS}{Operating System}
\newacronym{PAg}{PAg}{Per-address Adaptive branch prediction using one Global PHT}
\newacronym{PAs}{PAs}{Per-address Adaptive branch prediction using per-Set PHT}
\newacronym{PC}{PC}{Program Counter}
\newacronym[longplural={Pointer-Chasing Engines}]{PCE}{PCE}{Pointer-Chasing Engine}
\newacronym{PCM}{PCM}{Performance Counter Monitor}
\newacronym{PIM}{PIM}{processing-in-memory}
\newacronym[longplural={Pattern History Tables}]{PHT}{PHT}{Pattern History Table}
\newacronym{RAPL}{RAPL}{Running Average Power Limit}
\newacronym{RAS}{RAS}{Row Address Strobe}
\newacronym{RAT}{RAT}{Registers Alias Table}
\newacronym{RC}{RC}{Row Cycle}
\newacronym{RCD}{RCD}{RAS to CAS Delay}
\newacronym[longplural={Row-based Database Systems}]{RDBS}{RDBS}{Row-based Database System}
\newacronym{ROB}{ROB}{Reorder Buffer}
\newacronym{RRD}{RRD}{Row to Row activation Delay}
\newacronym{DNN}{DNN}{Deep Neural Network}
\newacronym{ANN}{ANN}{Artificial Neural Network}
\newacronym{PnM}{PnM}{processing-near-memory}
\newacronym{PuM}{PuM}{processing-using-memory}

\newacronym{RP}{RP}{Row Precharge}
\newacronym{RTL}{RTL}{Register Transfer Level}
\newacronym{RTP}{RTP}{Read To Precharge}
\newacronym{RVU}{RVU}{Reconfigurable Vector Unit}
\newacronym{SDP}{SDP}{Skewed Dead-Block Predictor}
\newacronym{SESC}{SESC}{Superescalar Simulator}
\newacronym{SSE}{SSE}{Streaming SIMD Extensions}
\newacronym{SFP}{SFP}{Spatial Footprint Predictor}
\newacronym{SiNUCA}{SiNUCA}{Simulator of Non-Uniform Cache Architectures}
\newacronym{SMT}{SMT}{Simultaneous Multi-Threading}
\newacronym{SIMD}{SIMD}{Single Instruction Multiple Data}
\newacronym{SPEC}{SPEC}{Standard Performance Evaluation Corporation}
\newacronym{SoC}{SoC}{System-on-Chip}
\newacronym{SPP}{SPP}{Spatial Pattern Predictor}
\newacronym{SRAM}{SRAM}{Static Random Access Memory}
\newacronym{SSOR}{SSOR}{Symmetric Successive Over-Relaxation}
\newacronym{SSV}{SSV}{Search Set Vector}
\newacronym{TLB}{TLB}{Translation Look-aside Buffer}
\newacronym{TLP}{TLP}{Thread Level Parallelism}
\newacronym[longplural={Through-Silicon Vias}]{TSV}{TSV}{Through-Silicon Via}
\newacronym{VWQ}{VWQ}{Virtual Write Queue}
\newacronym{WTR}{WTR}{Write Recovery time}
\newacronym{WR}{WR}{Write To Read delay time}

\usepackage{footnote}
\usepackage{multicol}
\usepackage{graphicx}     
\usepackage{textcomp}
\usepackage{xspace}
\usepackage[normalem]{ulem}
\usepackage[labelfont=bf, textfont=bf, skip=2pt]{caption}
\usepackage[english]{babel}
\usepackage{booktabs}
\usepackage{multirow}
\usepackage{ltablex}
\usepackage{soul}
\usepackage{makecell}
\usepackage[binary-units=true]{siunitx}
\usepackage{xargs} 
\usepackage{xcolor, colortbl}
\usepackage[multiple]{footmisc}
\usepackage{marginnote}
\usepackage{clipboard}
\usepackage{tikz}
\usepackage{subcaption}
\usepackage{mathtools}
\usepackage[compact]{titlesec}
\usepackage{wrapfig}
\usepackage{graphbox}
\usepackage{cuted,capt-of}
\usepackage[noadjust]{cite}
\usepackage{setspace}
\usepackage{todonotes}
\usepackage[colorlinks,urlcolor=blue,linkcolor=blue,citecolor=blue]{hyperref}
\usepackage[many]{tcolorbox}
\usepackage{amsmath,amssymb,amsthm}
\usepackage{filecontents}
\usepackage[italic]{mathastext}
\usepackage{pifont}
\let\labelindent\relax
\usepackage{enumitem}

\usepackage[pscoord]{eso-pic}

\newcommand{\placetextbox}[3]{
  \setbox0=\hbox{#3}
  \AddToShipoutPictureFG*{
    \put(\LenToUnit{#1\paperwidth},\LenToUnit{#2\paperheight}){\vtop{{\null}\makebox[0pt][c]{#3}}}%
  }%
}%

\jvol{XX}
\jnum{XX}
\paper{8}
\jmonth{}
\jname{}
\pubyear{}

\newcommand{\circled}[1]{\tikz[baseline=(char.base)]{\node[shape=circle,draw,inner sep=0pt,fill=black, text=white] (char) {#1};}}

\newtcolorbox{NewBox}[1]{%
  floatplacement={#1}, width=\columnwidth,
  colframe=gray!10!black,colback=orange!10!white,boxrule=1pt,arc=.2em,boxsep=-1.6mm,
  beforeafter skip balanced=0pt,
  }
  
\definecolor{Blue}{rgb}{0.06, 0.14, 0.57}

\newcommand{\gfcri}[1]{\textcolor{black}{#1}}
\newcommand{\gfcrii}[1]{\textcolor{black}{#1}}

\newcommand{\sg}[1]{\textcolor{black}{#1}}

\newcommand{\li}{\gfcri{(1)}}
\newcommand{\lii}{\gfcri{(2)}}
\newcommand{\liii}{\gfcri{(3)}}
\newcommand{\liv}{\gfcri{(4)}}
\newcommand{\lv}{\gfcri{(5)}}
\newcommand{\lvi}{\gfcri{(6)}}

\definecolor{lightblue}{rgb}{0.980, 0.956, 0.623}
\sethlcolor{lightblue}

\newcommand{\titleShort}{Mensa\xspace}
\newcommand{\exampleDesign}{\titleShort-G\xspace}
\newcommand{\accelA}{Pascal\xspace}
\newcommand{\accelB}{Pavlov\xspace}
\newcommand{\accelC}{Jacquard\xspace}

\newcommand{\base}{\emph{Baseline}\xspace}
\newcommand{\basehb}{\emph{Base+HB}\xspace}
\newcommand{\mensag}{\emph{Mensa-G}\xspace}

\DeclareSIUnit{\flop}{FLOP}

\newcommand{\mech}{{SIMDRAM}\xspace} 
\newcommand\edgemodels{edge NN models\xspace}

\definecolor{dollarbill}{rgb}{0.52, 0.73, 0.4}

\newcommandx{\unsure}[2][1=]{\todo[linecolor=red,backgroundcolor=red!25,bordercolor=red,#1, size=\tiny]{#2}}
\newcommandx{\change}[2][1=]{\todo[linecolor=blue,backgroundcolor=blue!25,bordercolor=blue,#1,size=\tiny]{#2}}
\newcommandx{\feedback}[2][1=]{\todo[linecolor=yellow,backgroundcolor=yellow!25,bordercolor=yellow,#1]{#2}}
\newcommandx{\improvement}[2][1=]{\todo[linecolor=Plum,backgroundcolor=Plum!25,bordercolor=Plum,#1]{#2}}
\newcommandx{\thiswillnotshow}[2][1=]{\todo[disable,#1]{#2}}
\newcommandx{\completedRevision}[2][1=]{\todo[disable,backgroundcolor=red,#1]{#2}}
\newcommandx{\dataSource}[2][1=]{\todo[disable,backgroundcolor=red,#1]{#2}}
\newcommandx{\info}[2][1=]{\todo[linecolor=dollarbill,backgroundcolor=dollarbill!25,bordercolor=dollarbill,#1, size=\tiny]{#2}}

\newif\ifrevision
\revisiontrue

\ifrevision 
    \newcommand{\juan}[1]{\textcolor{black}{#1}}
    \newcommand{\gf}[1]{\textcolor{black}{#1}}
    \newcommand{\gfi}[1]{\textcolor{black}{#1}}
    \newcommand{\gfii}[1]{\textcolor{black}{#1}}
    
    \newcommand{\gfrev}[1]{\textcolor{black}{#1}}

    \renewcommand\hl[1]{#1} 
\else
    \newcommand{\juan}[1]{\textcolor{black}{#1}}

    \newcommand{\gf}[1]{\textcolor{black}{#1}}
    \newcommand{\gfi}[1]{\textcolor{black}{#1}}
    \newcommand{\gfii}[1]{\textcolor{black}{#1}}
    
    \newcommand{\gfrev}[1]{\textcolor{black}{#1}}
    
\fi 

\newcommand\pimdef{\cite{ghose.ibmjrd19, mutlu2020modern,deoliveira2021IEEE,pim-book,mutlu2019processing,mutlu2019enabling,mutlu2015research,mutlu2013memory,loh2013processing,Near-Data,stone1970logic,Miss_Mem_Wall_1996}\xspace}

\newcommand\pnmtwod{\cite{farmahini2015nda,babarinsa2015jafar,devaux2019true,ghiasi2022genstore,gomez2021benchmarkingcut,gomezluna2021benchmarking,gomez2022benchmarking,syncron,singh2020nero,skhynixpim,ke2021near,giannoula2022sparsep,shin2018mcdram,cho2020mcdram,denzler2021casper,asghari2016chameleon,IRAM_Micro_1997,C_RAM_1999,CASES_MVX,Xi_2015,sun2021abc,matam2019graphssd,gokhale1995processing,hall1999mapping,MEMSYS_MVX,lockerman2020livia}\xspace}

\newcommand\pnmthreed{\cite{ahn2015scalable,nai2017graphpim,boroumand2018google,lazypim, top-pim, gao2016hrl, kim2018grim, drumond2017mondrian, RVU, NIM, PEI, gao2017tetris,Kim2016,gu2016leveraging, boroumand2019conda, hsieh2016transparent, cali2020genasm, NDC_ISPASS_2014,pattnaik2016scheduling,akin2015data,hsieh2016accelerating,lee2015bssync,boroumand2021mitigating,boroumand2021google,boroumand2022polynesia,boroumand2021polynesia,amiraliphd,besta2021sisa,fernandez2020natsa,singh2019napel,kwon202125,lee2021hardware,niu2022184qps,Sparse_MM_LiM,azarkhish2016logic,azarkhish2018neurostream,guo20143d,de2018design,akin2014hamlet,huang2020heterogeneous,dai2018graphh,liu2018processing,tsai:micro:2018:ams,gu2020ipim,DRAMA_CAL_2014,Asghari-Moghaddam_2016,huang2019active,kersey2017lightweight,li2019pims,kim2017grim,boroumand2017lazypim,zhuo2019graphq,zhang2018graphp,lim2017triple,smc_sim,HIVE,jang2019charon,IBM_ActiveCube,hadidi2017cairo,santos2018processing}\xspace}

\newcommand\pum{\cite{Chi2016, Shafiee2016, seshadri2017ambit, seshadri2019dram, li2017drisa, seshadri2013rowclone, seshadri2016processing, deng2018dracc, xin2020elp2im, song2018graphr, song2017pipelayer,gao2019computedram, eckert2018neural, aga2017compute,dualitycache,besta2021sisa,seshadri2016buddy,seshadri.bookchapter17,seshadri2018rowclone,seshadri2015fast,li2016pinatubo,ferreira2021pluto,ferreira2022pluto,imani2019floatpim,he2020sparse,flashcosmos,truong2022adapting,truong2021racer,olgun2021quactrng,kim2019d,kim2018dram,bostanci2022dr,olgun2022pidram,ali2019memory,angizi2019graphide,li2018scope,subramaniyan2017parallel,zha2020hyper,fujiki2018memory,orosa2021codic,sharad2013ultra,rezaei2020nom}}

\newcommand\drampimshort{\cite{ahn2015scalable,akin2014hamlet,akin2015data,ali2019memory,amiraliphd,angizi2019graphide,Asghari-Moghaddam_2016,asghari2016chameleon,azarkhish2016logic,azarkhish2018neurostream,babarinsa2015jafar,besta2021sisa,boroumand2017lazypim,boroumand2018google, boroumand2019conda,boroumand2021google,boroumand2021mitigating,boroumand2021polynesia,boroumand2022polynesia,bostanci2022dr, cali2020genasm}}

\newcommand\srampim{\cite{aga2017compute,denzler2021casper,dualitycache,eckert2018neural,gokhale1995processing,lockerman2020livia}}

\newcommand\nvmpim{\cite{fujiki2018memory,imani2019floatpim,Kim2016,li2016pinatubo,Shafiee2016,song2017pipelayer,song2018graphr,truong2021racer,zha2020hyper,truong2022adapting,sharad2013ultra}}

\newcommand\upmem{\cite{upmem,upmem2018,devaux2019true,gomez2022benchmarking,gomezluna2021benchmarking,gomez2021benchmarkingcut}}

\newcommand\ambit{\cite{seshadri2017ambit,seshadri2019dram,seshadri2015fast,seshadri.bookchapter17,seshadri2016buddy,seshadri2016processing}}

\begin{document}
\bstctlcite{IEEEexample:BSTcontrol}

\placetextbox{0.55}{0.87}{\textsf{\emph{This is an extended and updated version of a paper published in}}}%
\placetextbox{0.55}{0.85}{\textsf{\emph{IEEE Micro, pp. 1--14, 29 Aug. 2022}}}%
\placetextbox{0.55}{0.83}{\textsf{\emph{\url{https://doi.org/10.1109/MM.2022.3202350}}}}%

\title{\scalebox{0.91}{Accelerating Neural Network Inference}\\ \scalebox{0.91}{with Processing-in-DRAM:} \\ \scalebox{0.91}{From the Edge to the Cloud}\vspace{-10pt}}

\author{Geraldo F. Oliveira}
\affil{ETH Zürich}

\author{Juan Gómez-Luna}
\affil{ETH Zürich}

\author{Saugata Ghose}
\affil{University of Illinois Urbana\gfcri{-}Champaign}

\author{Amirali Boroumand}
\affil{Google}

\author{Onur Mutlu}
\affil{ETH Zürich\vspace{-15pt}
}

\begin{abstract}
    \Copy{R25}{\glspl{NN} are growing in importance and complexity. \gfcri{A neural network}'s performance (and energy efficiency) can be bound either by computation or memory resources. The \gls{PIM} paradigm, where computation is placed near or within memory arrays, is a viable solution to accelerate memory-bound \glspl{NN}. However, \gls{PIM} architectures vary in form, where different \gls{PIM} approaches lead to different trade-offs. Our goal is to analyze, discuss, and contrast DRAM-based \gls{PIM} architectures for \gls{NN} performance and energy efficiency. To do so, we analyze three state-of-the-art \gls{PIM} architectures: 
(1)~UPMEM, which integrates processors and DRAM arrays into a single 2D chip\gfcri{;}
(2)~Mensa, a 3D-\gfcri{stack-}based \gls{PIM} architecture tailored for edge devices\gfcri{;} and
(3)~SIMDRAM, which uses the analog principles of DRAM to execute bit-serial operations.}
Our analysis reveals that \gls{PIM} greatly benefits memory-bound \glspl{NN}:
\gfcri{(1)~}UPMEM provides 23$\times$ the performance of a high-end GPU when the GPU requires memory oversubscription for a \gfcri{general matrix\sg{--}vector multiplication} kernel\gfcri{;} 
\gfcri{(2)~}Mensa improves energy efficiency and throughput by 3.0$\times$ and 3.1$\times$ over the \gfcri{Google} Edge TPU for 24 \gfcri{Google} edge \gls{NN} models\gfcri{;} and
\gfcri{(3)~}SIMDRAM outperforms a CPU/GPU by 16.7$\times$/1.4$\times$ for three binary \glspl{NN}. 
\sg{We conclude that the ideal PIM architecture for NN models depends on a model's distinct attributes, due to the inherent architectural design choices.}

\vspace{-5pt}
\end{abstract}

\maketitle

\section{Introduction}
\label{sec:introduction}

\chapterinitial{N}eural networks (NNs)~\gfcrii{\cite{rosenblatt1958perceptron,ivakhnenko1965,yegnanarayana2009artificial}} are becoming increasingly important for \gf{many daily activities}, from routine \gfcri{tasks such as} commute traffic estimation~\gfcrii{\cite{oliveira2016computer}} to critical \gfcri{tasks such as} as medical diagnosis~\gfcrii{\cite{al2011artificial}}. 
NN algorithms are rapidly evolving, which has led to many different types of NN models~\gfcrii{\cite{sze2017efficient,fukushima.biologicalcybernetics1980, lecun.cognitiva1985, rumelhart.nature1986, lecun.nature2015, simonyan2015very,gu2018recent,lecun1989handwritten,lecun1998gradient,russakovsky2015imagenet,zeiler2014visualizing,szegedy2015going,he.cvpr2016,hochreiter.neco1997, gers.icann1999, greff.tnnls2017,lstm-google,graves2013generating,google-translation,cho2014learning, lrcn, karpathy2015deep,ranzato2014video,srivastava2015unsupervised,sutskever2014sequence,xu2015show,xingjian2015convolutional,gru,kanai2017preventing,cho2014properties,graves.icmlworkshop2012, he.icassp2019,liang.cvpr2015, pinheiro.icml2014,rcnn-google,rastegari2016xnor}}. Such NN algorithms behave as \juan{either} \emph{compute-bound} workloads (i.e., when computing resources are the main performance bottleneck) or as \emph{memory-bound} workloads (i.e., when memory resources are the main performance bottleneck) depending on the NN model~\gfcrii{\cite{vanhoucke2011improving,jain2018architectural,adolf2016fathom,boroumand2021google,tpu,reddi2020mlperf,wang2020neural,wang2020systematic,wang2019benchmarking,gupta2019architectural}}.

\gfi{A} common approach to improve \gfcri{the} performance \gfi{of compute-bound NNs} is to design compute-centric accelerators~\gfcrii{\cite{sze2017efficient, edge-tpu,vanhoucke2011improving,chen2018tvm,dnpu,eyeriss,tpu,wang2020neural,eyerissv2,scnn,han2016eie,kim2018deeptrain,tangram,han2017ese,lrcn-fpga}}, where \gfcri{a} significant portion of hardware resources are \gfcri{dedicated to} the processing elements of the \gf{accelerator}. Compute-centric NN accelerators \gfcri{aim to speed up two main operations} that \gfii{many} NN models rely on~\gfcrii{\cite{adolf2016fathom}}: 
multiply-and-accumulate (MAC)\gf{, largely used for convolutions\gfcri{;} and matrix-vector multiplications (MVMs)}. \Copy{E1A1}{Similarly, various approaches to improve \gfcri{the} performance of memory-bound NNs have been proposed~\cite{boroumand2021google,gao2017tetris,boroumand2018google,imani2019floatpim,koppula2019eden,liu2018processing,min2019neuralhmc,cho2020mcdram,Shafiee2016,eckert2018neural,Chi2016,peemen2013memory}. One such approach is 
processing-in-memory (PIM)~\gfcrii{\pimdef}, where computation is 
\li~placed  near \gf{2D~\gfcrii{\pnmtwod} or 3D~\gfcrii{\pnmthreed} memory arrays (processing-near-memory, \gfcri{PNM}) or \lii~performed using the memory arrays themselves (processing-using-memory, \gfcri{PUM})~\gfcrii{\pum}}.}\footnote{\label{R1/1}\Copy{E1A2}{\gfrev{\hl{There is no consensus on the terminology related to processing-in-memory in the current literature. However, we borrow the terms ``processing-near-memory'' (\gfcri{PNM}) and ``processing-using-memory'' (\gfcri{PUM}) from~\mbox{\cite{mutlu2020modern}}, which clearly defines these two terms and demonstrates that existing PIM \gfcri{techniques} fall into one of these two categories.}}}\label{foot:r1/1}} \Copy{E1A3}{\gfii{\gfcri{PNM} and \gfcri{PUM} can be implemented using different memory technologies~\cite{mutlu2020modern}, including SRAM~\gfcrii{\srampim}, DRAM~\gfcrii{\drampimshort}, and resistive RAM (\gfcri{ReRAM})~\gfcrii{\nvmpim}. \gfcri{Independent of} the memory technology,} PIM allows NNs to \gfrev{\hl{enjoy higher}} memory bandwidth, shorter memory access latency, and lower energy per bit~\cite{mutlu2020modern}. 
}

\Copy{E1B}{Our \emph{goal} is to analyze, discuss, and contrast the benefits and drawbacks of different PIM architectures for NN performance and energy efficiency \gfrev{\hl{in}} different computing environments (i.e., edge devices and cloud servers). \gfii{In this work, we focus on DRAM-based PIM,} \gfcri{and} NN inference in three different \gfcri{DRAM-based} PIM \gfii{architectures}:
\gfrev{\hl{
\li~UPMEM~\gfcrii{\upmem}, a large-scale 2D-based \gfcri{PNM} architecture, which integrates several simple general-purpose \gfcri{processing} cores near DRAM arrays \gfcri{(inside a DRAM chip)};
\lii~Mensa~\mbox{\cite{boroumand2021google}}, a heterogeneous \gfcri{edge} NN inference accelerator, which uses 3D-\gfcri{stack}-based \gfcri{PNM} to speed up memory-bound NN models; and
\liii~SIMDRAM~\mbox{\cite{hajinazarsimdram}}, a \gfcri{PUM} architecture, which uses the analog properties of DRAM to execute bit-serial operations.}} \gfrev{\hl{These three \emph{state-of-the-art} PIM architectures broadly cover the design space for PIM designs since 
\li~UPMEM represents the design of 2D-based \gfcri{PNM} systems,
\lii~Mensa represents the design of a \gfcri{PNM} system that leverages 3D-stacked integration to add specialized hardware units near DRAM, and
\liii~SIMDRAM represents the design of \gfcri{PUM} systems.}}}

\Copy{E1C}{We draw three key takeaways from our analysis. First, the evaluated PIM architectures provide significant performance improvement and energy savings for NN inference when compared against compute-centric architectures (i.e., CPU\gf{s}\gfi{,} GPU\gf{s}\gfi{, and \gfcri{compute-centric edge TPUs}}). Second, there is \emph{no} holy grail PIM architecture that can broadly accelerate \emph{all} different types of NN models. This happens because different PIM architectures come with different trade-offs \gf{(e.g., limited memory capacity, high cost, high complexity of mapping workloads to the PIM substrate, \gfcri{lack of} support for key operations in NN inference).} \gfrev{\hl{However, regardless of the underlying operating principles of our three state-of-the-art PIM architectures, they all can alleviate the memory bottlenecks our evaluated NN models suffer from by leveraging the internal memory bandwidth and abundant parallelism available inside DRAM.}}

Third, there is a need for programming models and frameworks that can unify the benefits of the different PIM architectures into a single heterogeneous system. Such solutions need to map, schedule, and control the execution of NN models \gfcri{on} the most appropriate PIM architecture. Thus, \gfcri{they can enable a}  workload to leverage the benefits of PIM while avoiding particular drawbacks related to a given PIM \gfcri{architecture}. We conclude that PIM is a promising solution to improve the performance and energy efficiency of several NN models.} 

We make the following key contributions:
\begin{itemize}
\item We investigate how three PIM architectures can be used to improve \gfcri{the} performance and energy efficiency of different NN models. 

\item \gfrev{\hl{We observe that the fundamental properties of the target PIM architecture define the most suitable type of NN model for the underlying \gfcri{architecture}.}}

\end{itemize}

\section{Drawbacks of Compute-Centric NN Accelerators}
\label{sec:motivation}
\label{sec:motiv}
\label{sec:motiv:methodology}

We analyze the performance and energy of executing NN models using a commercial state-of-the-art compute-centric NN accelerator, called \sg{the Google Edge TPU}~\cite{edge-tpu}, as our baseline accelerator. The Edge TPU has a generic tiled architecture, similar to other state-of-the-art accelerators~\cite{gao2017tetris, sze2017efficient}. It includes a 64$\times$64 2D array of floating-point \gfrev{\hl{multiply-and-accumulate}} processing elements (PEs), where each PE has a small register file to hold intermediate results.  The accelerator has two large SRAM-based on-chip buffers to  hold model parameters \gfrev{\hl{(\SI{4}{\mega\byte} parameter buffer)}} and activations \gfrev{\hl{(\SI{2}{\mega\byte} activation buffer)}}. We analyze 24 Google edge \gfcri{NN} models, including convolutional neural networks (CNNs)~\gfcrii{\cite{fukushima.biologicalcybernetics1980, lecun.cognitiva1985, rumelhart.nature1986, lecun.nature2015, simonyan2015very}}, long short-term memory (LSTM) networks~\gfcrii{\cite{hochreiter.neco1997, gers.icann1999, greff.tnnls2017}}, transducers~\gfcrii{\cite{graves.icmlworkshop2012, he.icassp2019}}, and recurrent convolutional neural networks (RCNNs)~\gfcrii{\cite{liang.cvpr2015, pinheiro.icml2014, lrcn}}. Based on our analysis, we find that the accelerator suffers from three \gfcrii{key} shortcomings\gfcri{.}

\begin{figure}[ht]
    \vspace{-5pt}
    \centering
    \centering
    \includegraphics[width=\linewidth]{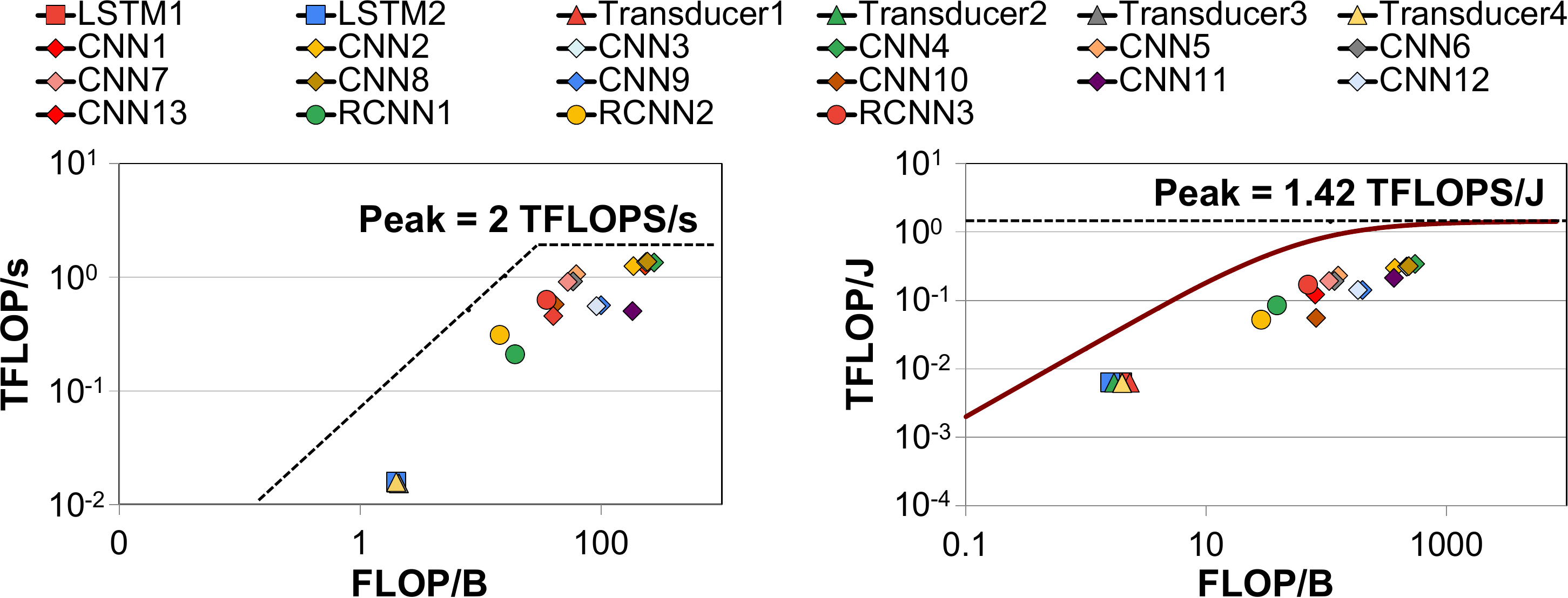}%
    \caption{Throughput roofline (left) and energy \gfcri{efficiency} roofline (right) for the Edge TPU across all Google edge NN models.} 
    \label{fig:roofline-throughput}
    \label{fig:roofline-energy}
    \vspace{-8pt}
\end{figure}

\textbf{1. The accelerator often suffers from extreme underutilization of the PEs.} The Edge TPU has a theoretical peak throughput of \SI{2}{\tera\flop\per\second}. However, the accelerator operates \gfcri{at a throughput} \emph{much} lower than \gfi{its} peak throughput during inference execution (75.6\% lower on average). Figure~\ref{fig:roofline-throughput} (left) shows the throughput roofline model~\gfcrii{\cite{williams2009roofline}} 
 for the Edge TPU, along with the measured throughput of all of our edge models. \gf{We observe that t}he PE utilization is low across all models. Transducer and LSTM models have the \gfcri{highest} underutilization, with both achieving less than 1\% of peak throughput. While CNN and RCNN models do somewhat better, they achieve only 40.7\% of peak utilization on average (with a minimum of only 10.2\% of peak utilization). We identify two main reasons why the actual throughput of the Edge TPU falls significantly \gfi{below its} peak.

First, while some layers have high parameter reuse (\SI{1200}{\flop\per\byte} ratio), other layers exhibit very low reuse (\SIrange{1}{64}{\flop\per\byte}) while at the same time having large parameter footprints (\SIrange{0.5}{5}{\mega\byte}). Layers with low reuse yet large \gfi{parameter} footprints are \emph{memory-bound} and often leave PEs idle, as the parameters incur long-latency cache misses to DRAM. 

Second, the Edge TPU does \emph{not} provide a custom dataflow optimized for each layer. From our analysis, we observe that layers both across and within models exhibit high variation in terms of data reuse patterns and computational intensity. This variation necessitates the need for \emph{different} dataflows for different layers, where each dataflow exposes a different set of reuse opportunities for parameters and activations. 

\begin{figure}[ht]
    \vspace{-5pt}
    \centering
        \centering
        \includegraphics[width=\linewidth]{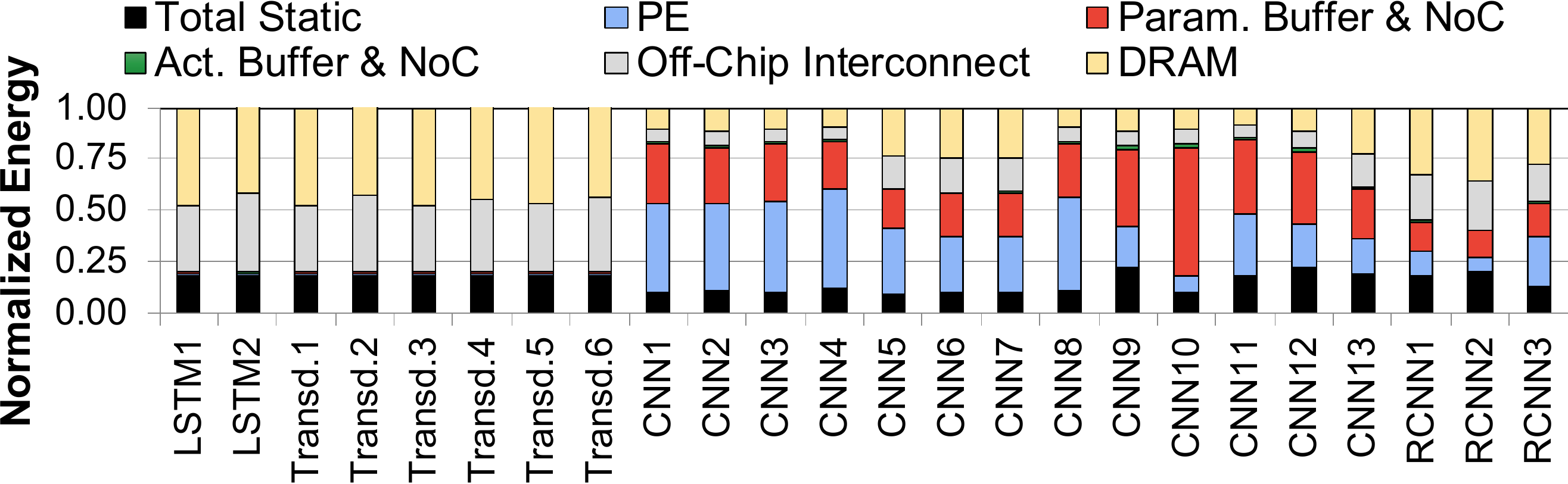}%
    \caption{Energy breakdown during inference.}
    \label{fig:energy-breakdown}
\vspace{-8pt}
\end{figure}

\textbf{2. \gfi{T}he Edge TPU operates far below its
theoretical maximum energy efficiency.} 
Figure~\ref{fig:roofline-energy} (right) shows the energy \gfcri{efficiency} roofline~\gfcrii{\cite{energy-roofline-model}} for the Edge TPU, along with the \gfi{energy} efficiency achieved for each model. 
\gfcrii{On} average across all models, the Edge TPU achieves only 37.2\% of its maximum possible energy efficiency. The energy efficiency is  particularly low (33.8\% of the maximum) for LSTM and Transducer models, but even the best CNN model achieves only 50.7\% of the maximum \gfi{energy} efficiency. 
To understand the accelerator's low energy efficiency, we analyze (in Figure~\ref{fig:energy-breakdown}) the energy breakdown during inference execution across different models.  We make three key observations.

First, the on-chip buffers (the activation buffer and the parameter buffer) account for a significant portion of both static and dynamic energy across all models. 
Second, averaged across all models, the Edge TPU spends 50.3\% of its total energy on off-chip memory accesses. Third, for LSTMs and Transducers, the Edge TPU spends approximately three quarters of its total energy on DRAM accesses. This is because while the buffers consume a significant amount of area in the Edge TPU (79.4\% of the total area), they are ineffective at reducing off-chip memory accesses. 

\textbf{3. The accelerator's memory system design is neither effective nor efficient.} \gfi{T}he Edge TPU memory system\gfi{,} which includes both on-chip buffers and off-chip memory\gfi{,} is highly inefficient, and results in significant energy consumption.
 
\noindent \textbf{Key Takeaways.} Our analysis provides two key insights:
(1)~there is significant variation in terms of layer characteristics \emph{across} and 
\emph{within} edge NN models, where models vary from being compute-bound and memory-bound; 
(2)~the monolithic design of state-of-the-art NN accelerators mainly caters for compute-bound NN models, which leads to high underutilization and low energy efficiency of memory-bound NN models.

\section{PIM Architectures for NN Inference}

The \emph{goal} of our work is \gf{to} explor\gf{e} \sg{how} different state-of-the-art PIM architectures\footnote{\gfii{We refer the reader to the sidebar for a
brief discussion of several other related works.}} \sg{overcome} the memory bottlenecks caused by data movement in NN inference \gf{in different environments (i.e., from edge devices to cloud servers).}

\Copy{E2P2A}{We conduct our analysis in three steps. \gfrev{\hl{First, we evaluate the performance of a key primitive in NN inference (i.e., general matrix\sg{--}vector multiplication, GEMV) on the first commercial server-based PIM architecture (UPMEM~\gfcrii{\upmem}), which is a 2D general-purpose \gfcri{PNM} architecture. 
Second, we design specialized NN inference accelerators for 24 different edge NN models in a 3D-stacked memory using the Mensa framework~\mbox{\cite{boroumand2021google}}. Since Mensa relies on the 3D integration of logic and memory, the micro-architecture for the PIM accelerators can consist of an extensive range of processing elements, including costly floating-point MAC units.
Third, we map three different binary neural networks (BNNs)~\gfcrii{\cite{rastegari2016xnor,lin2017towards,xiang2017binary,qin2020bipointnet,chen2021bnn}} to a \gfcri{PUM} substrate called SIMDRAM~\mbox{\cite{hajinazarsimdram}}, which is a promising solution for both edge-based and cloud-based systems since it operates using standard 2D DDRx DRAM chips. We conclude by drawing key takeaways from our analyses, and based on them, we provide some guidelines for future PIM systems targeting NN inference.}}}

\Copy{E2P2B}{\gfrev{\hl{These three \emph{state-of-the-art} PIM architectures broadly cover the design space for general-purpose 2D \gfcri{PNM} (UPMEM), specialized 3D \gfcri{PNM} (Mensa), and DRAM-based \gfcri{PUM} (SIMDRAM) proposals. Regardless of the underlying operating principles of our three state-of-the-art PIM architectures, they all leverage \gfcri{DRAM's} abundant internal parallelism and \gfcri{internal} memory bandwidth to alleviate memory bottlenecks.}}} 

\section{\juan{NN Inference on General-Purpose 2D-Based Processing-near-Memory}}
\label{sec:upmem}

\Copy{R1/11A}{\juan{We conduct a preliminary analysis of the potential of \gf{the first \sg{commercially-available} server-based PIM system, called UPMEM~\gfcrii{\upmem}, for NN inference}. Many NN inference workloads (e.g., CNN, LSTM) use matrix\sg{--}vector multiplication (GEMV) as a key operation.  GEMV is characterized as memory-bound in \gf{compute}-centric architectures (i.e., CPU, GPU)~\cite{deoliveira2021IEEE,gomez2022benchmarking}. 
As a result, GEMV is a suitable candidate to be mapped onto the UPMEM architecture.}}\footnote{\label{R1/12}\Copy{R1/11B}{\gfrev{\hl{We show GEMV performance in our UPMEM experiments as a preliminary evaluation of the benefits the UPMEM system can provide for NN inference. Implementing an end-to-end NN inference model using the UPMEM system requires non-trivial effort, which we leave for future work. }}}\label{ft:R1/11B}}\Copy{R1/11C}{ 
We compare quantitatively and qualitatively our PIM implementation of GEMV to its state-of-the-art GPU counterpart.}

\subsection{\juan{UPMEM PIM Architecture}}

\begin{figure}[ht]
   \vspace{-5pt}
    \centering
    \includegraphics[width=\linewidth]{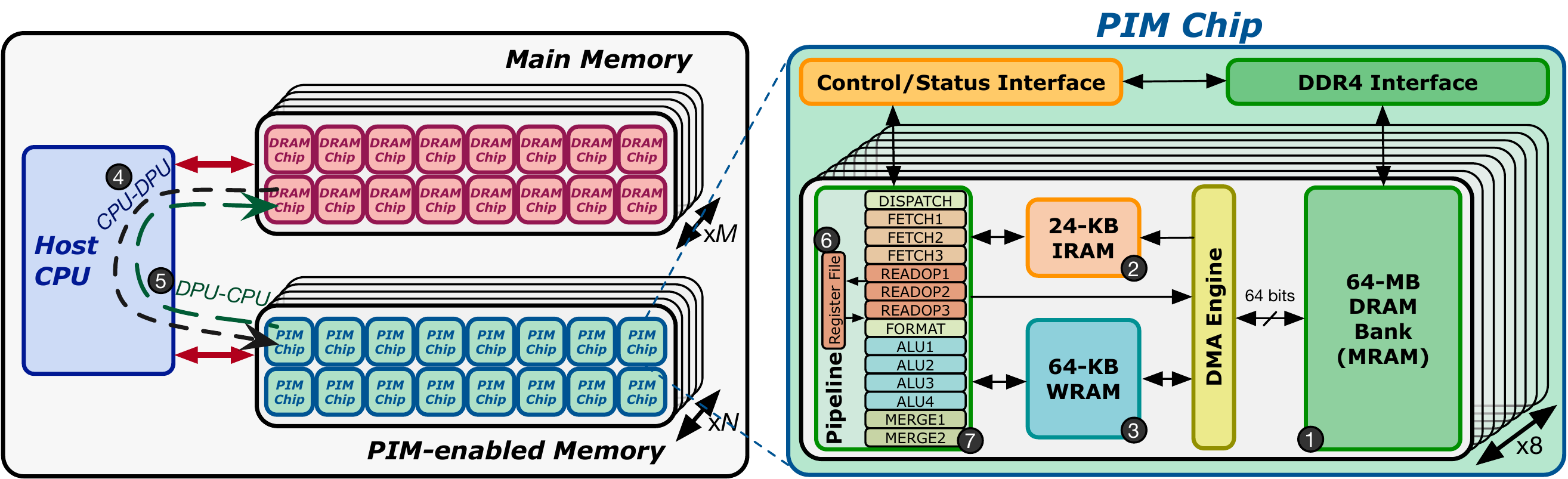}
    \caption{UPMEM-based PIM system~\cite{gomez2022benchmarking, devaux2019true}.}
    \label{fig:scheme}
    \vspace{-10pt}
\end{figure}

\juan{Figure~\ref{fig:scheme} (left) depicts \sg{an} UPMEM-based PIM system with (1)~a \emph{host} CPU, (2)~standard main memory (DRAM modules), and (3)~PIM-enabled memory (UPMEM modules)~\cite{ devaux2019true}. \sg{An} UPMEM module is a standard DDR4-2400 DIMM (module) with 16 PIM chips.} 

\label{R2/4C}\Copy{R24}{\juan{Inside each UPMEM PIM chip (Figure~\ref{fig:scheme}, right), there are 8 small general-purpose in-order processors, called \emph{DPUs}. DPUs are fine-grained multithreaded. Each DPU has exclusive access to (1) a \SI{64}{\mega\byte} DRAM bank, called \emph{MRAM} \circled{1}; (2)~a \SI{24}{\kilo\byte} instruction memory, called \emph{IRAM}  \circled{2}; and (3)~a \SI{64}{\kilo\byte} scratchpad memory, called \emph{WRAM} \circled{3}. MRAM is accessible by the host CPU (Figure~\ref{fig:scheme}, left) for \emph{copying} input data (from main memory to MRAM) \circled{4} and \emph{retrieving} results (from MRAM to main memory) \circled{5}. \gfrev{\hl{In UPMEM, data is explicitly copied from \sg{a} CPU to DPUs, and from MRAM to WRAM using CPU instructions. The UPMEM runtime system transposes DPU data using software libraries so that \gfcri{a 64-bit} word is placed inside a single DPU. There is no communication path across DPUs: all inter-DPU communication happens through the CPU.}} In current UPMEM-based PIM \gfrev{\hl{systems}}, the maximum number of UPMEM DIMMs is 20. \gfcrii{Thus}, the PIM system contains up to 2,560 DPUs which amounts to \SI{160}{\giga\byte} of PIM-capable memory.}}

\subsection{Evaluation Methodology}

\label{pg:E3P3}\Copy{E3P3}{\juan{We evaluate GEMV on a \gfrev{\hl{real}} UPMEM-based PIM \gf{cloud} system with 2048 DPUs and on a state-of-the-art NVIDIA A100 GPU~\gfcrii{\cite{a100,choquette2020nvidia}}. 
We use an open-source implementation of GEMV for \gf{UPMEM}~\gfcrii{\cite{gomez2022benchmarking,gomezluna2021benchmarking,gomez2021benchmarkingcut} and the cuBLAS library~\gfcrii{\cite{cublas}} for \sg{the} GPU.}} \juan{DPUs run at \SI{428}{\mega\hertz}. The maximum aggregated \gfcri{memory} bandwidth for 2048~DPUs is \SI{1.7}{\tera\byte\per\second} for a \gfcri{memory} capacity of \SI{128}{\giga\byte}. 
The A100 GPU runs at \SI{1.4}{\giga\hertz}. The A100 GPU features \SI{40}{\giga\byte} \sg{of} HBM2 memory~\gfcrii{\cite{hbm2}} with a bandwidth of \SI{1.5}{\tera\byte\per\second}.}}

\juan{We perform two experiments. 
First, we evaluate performance and scalability characteristics of GEMV on the \gf{UPMEM} system. 
Second, we compare the performance of GEMV on 2048 DPUs with that on the GPU. Since the amount of memory of the GPU is relatively small, we evaluate one version with regular memory allocation and another version with unified memory allocation, which allows memory oversubscription.}

\subsection{Results}
\begin{figure}[ht]
    \vspace{-5pt}
    \centering
    \includegraphics[width=\linewidth]{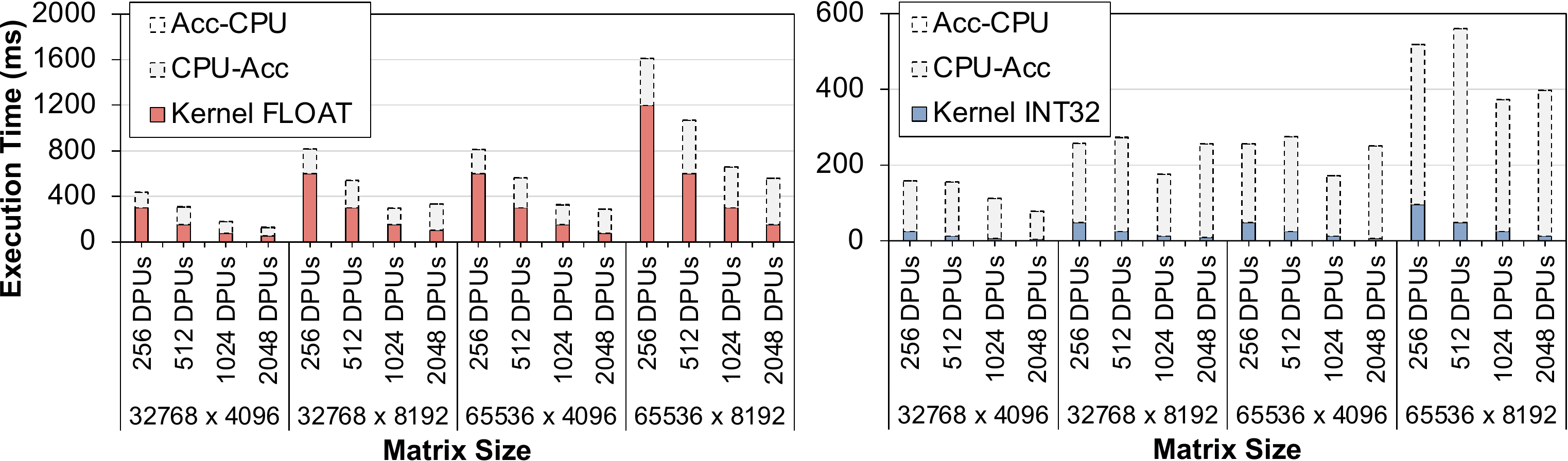}
    \caption{Execution time (ms) of GEMV for 32-bit floating-point \gf{(left)} and 32-bit integer \gf{(right)} on 256, 512, 1024, and 2048 DPUs.}
    \label{fig:gemv-dpus}
    \vspace{-8pt}
\end{figure}

\juan{Figure~\ref{fig:gemv-dpus} shows strong scaling results of GEMV for four different matrix sizes on the \gf{UPMEM} system. The matrices have 32-bit floating-point or integer values. The experiment uses 256, 512, 1024, and 2048 DPUs. Inside each DPU, \sg{we execute 16 software threads.}}

\juan{We make two observations. 
First, the execution with floating-point values takes one order of magnitude more time than the execution with integers. The reason is that the UPMEM DPUs do \emph{not} have native support for floating-point operations. This is a limitation that comes from the difficulty of integrating logic into DRAM chips~\gfcrii{\cite{kim1996assessing}}. As a result, floating-point operations are emulated~\cite{gomez2022benchmarking}. 
Second, the \gfi{kernel} execution \gfi{time} of GEMV (both floating-point and integer values) scales linearly on the \gf{UPMEM} system. The DPU kernel execution time reduces by nearly $2\times$ when we double the number of DPUs.}

\begin{figure}[ht]
    \centering
    \includegraphics[width=0.92\linewidth]{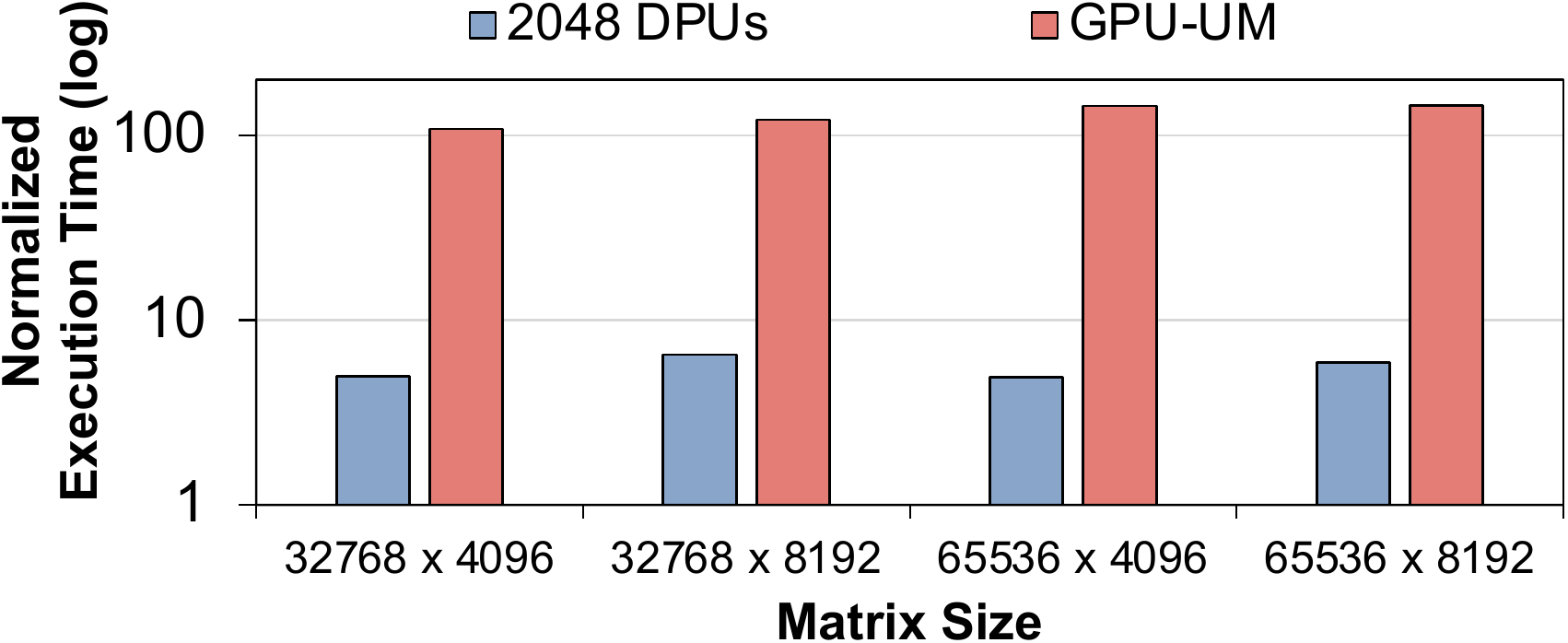}
    \caption{\gf{Normalized} execution time of GEMV for 32-bit integer values on 2048 DPUs and A100 GPU with unified memory (GPU-UM). \gf{Values are normalized to an  A100 GPU without unified memory.}}
    \label{fig:gemv-gpu-int32}
    \vspace{-5pt}
\end{figure}

\juan{Figure~\ref{fig:gemv-gpu-int32} shows the comparison of GEMV (integer values) on the PIM system with 2048~DPUs to the A100 GPU.} \gfcri{We observe that} the GEMV kernel execution time on the 2048-DPU \gf{UPMEM} system and on the GPU (without unified memory) is in the same order of magnitude, \gfcri{with} \sg{the} GPU \gfcri{being} 4--5$\times$ faster \gf{than the UPMEM system}. This is \gfcri{because}  
(1)~\gfcri{the peak compute throughput of the A100 GPU is orders of magnitude higher (TFLOPs vs. GOPs) even though} the \gf{UPMEM} system and the A100 GPU have similar memory bandwidth~\cite{gomez2022benchmarking}, and 
(2)~32-bit integer multiplication is emulated on the UPMEM DPUs (there is no 32-bit multiplier~\cite{devaux2019true}). 

\gfcri{There are three key takeaways} from this comparison\gfcri{. First, the GPU has \sg{significant} compute resources (including hardware multipliers), which leads to its higher performance. As future PIM systems integrate better (e.g., higher precision) multipliers and more compute resources \sg{as GPUs have done}, we expect their performance to get significantly higher.} 
\gfcri{Second,} \gfcri{the} GEMV \gfcri{kernel is} very memory-bound on GPU \gfcri{for large matrix sizes. 
In contrast, the UPMEM system balances GEMV's compute and memory requirements better.}
\gfcri{Third}, \gfcri{when employing a unified memory allocation in the GPU system, the GEMV kernel's execution time increases significantly. In this case, \emph{both} the 2048-\gf{DPU} system and the GPU without unified memory greatly outperform the GPU with unified memory.} Allowing memory oversubscription in the GPU has a \gfcri{large} performance cost \gfcri{due to overheads related to address translation and page swapping between the CPU and GPU memories}~\gfcrii{\cite{li2019framework,ausavarungnirun2017mosaic,ausavarungnirun2018mask}}. \gfcri{The UPMEM system avoids the need for a unified memory space, since it provides higher memory capacity and better scalability of computing throughput and memory size compared \sg{to} the GPU.}

\label{R1/13}\juan{We run one final experiment (not shown in the figure) where we compare GEMV on the \gf{UPMEM} system for 32-, 16-, and 8-bit integers. The DPU contains an 8-bit multiplier~\cite{devaux2019true}, which allows the 16-bit and the 8-bit versions \gf{to} be $1.75\times$ and $2.17\times$ faster than the 32-bit version, respectively. 
These results demonstrate the potential of the \gf{UPMEM} system for NN inference in cases where fixed-point computation and quantization are feasible, despite the inherent limitations derived from the integration of logic units inside the 2D DRAM chip.}

\section{NN Inference on Specialized 3D-Based Processing-near-Memory}
\label{sec:mensa}

Modern 3D-stacked memories~\gfcrii{\cite{HBM, HMC2,lee2016simultaneous}} such as High Bandwidth Memory~\gfcrii{\cite{HBM}} and the Hybrid Memory Cube~\cite{HMC2} include logic layers that have access to the high memory bandwidth available within a 3D-stacked memory chip~\cite{mutlu2020modern}. We evaluate the opportunities a 3D-\gfcri{stack}-based \gfcri{PNM} architecture provides for NN inference. To do so, we use~\titleShort~\sg{\cite{boroumand2021google}}, a new machine learning accelerator design framework that distributes the layers from an NN model across
a collection of smaller \gf{memory- and compute-centric} hardware accelerators that are  specialized \gfcri{to} the properties of different \gfcri{NN} layer types. By specializing each accelerator to a subset of \gfcri{NN} layers, \titleShort avoids the shortcomings of monolithic edge \gf{NN} accelerators, resulting in a highly-efficient and high-performance \gf{heterogeneous} accelerator with a much smaller area \gfcri{footprint}.

We employ \titleShort's methodology to design a specialized NN inference system for the \gfcri{Google} \gf{\edgemodels} discussed in our motivation. For each layer, we study the correlation between different characteristics. These characteristics include 
(1)~parameter reuse (\si{\flop\per\byte}), 
(2)~parameter footprint (\si{\mega\byte}), and
(3)~MAC intensity (defined by the number of MAC operations). Based on all of the layer characteristics that we analyze, we observe across all layers from all models that 97\% of the layers group into one of five layer families. \gf{We design different accelerators to cater to the properties of each layer \gfcri{family}.}

\label{R1/5}\gf{W}e identify that
(a)~layers in Families 1 and 2 share a high MAC intensity (20M--20M), 
small parameter footprint (\SIrange{1}{500}{\kilo\byte}), and moderate-to-high parameter reuse (\gf{81--20K~\SI{}{\flop\per\byte}}); while
(b)~layers in Families 3 and 4 share a low MAC intensity (0.1M--25M),
large parameter footprint (\SIrange{0.5}{18}{\mega\byte}), and low parameter reuse (\gf{\SIrange{1}{64}{\flop\per\byte}}).
Thus, we need \emph{at least} two different accelerator designs: one that caters to the compute-centric behavior of Families~1/2, and one that caters to the data-centric behavior of Families~3/4. Given our resource-constrained edge environment, we look to see if layers in Family~5, which have a low MAC intensity (similar to Families 3 and 4) but a relatively small parameter footprint (similar to Families 1 and 2), can benefit from one of these two approaches. We find that the low MAC intensity, along with the low parameter reuse by many Family~5 layers, allow the layers to benefit from many of the \gfcri{data}-centric optimizations that benefit Families 3 and 4, so we study \gfcri{Families~3/4/5} collectively as we design the data-centric accelerators.

\begin{figure}[ht]
    \centering
    \includegraphics[width=\linewidth]{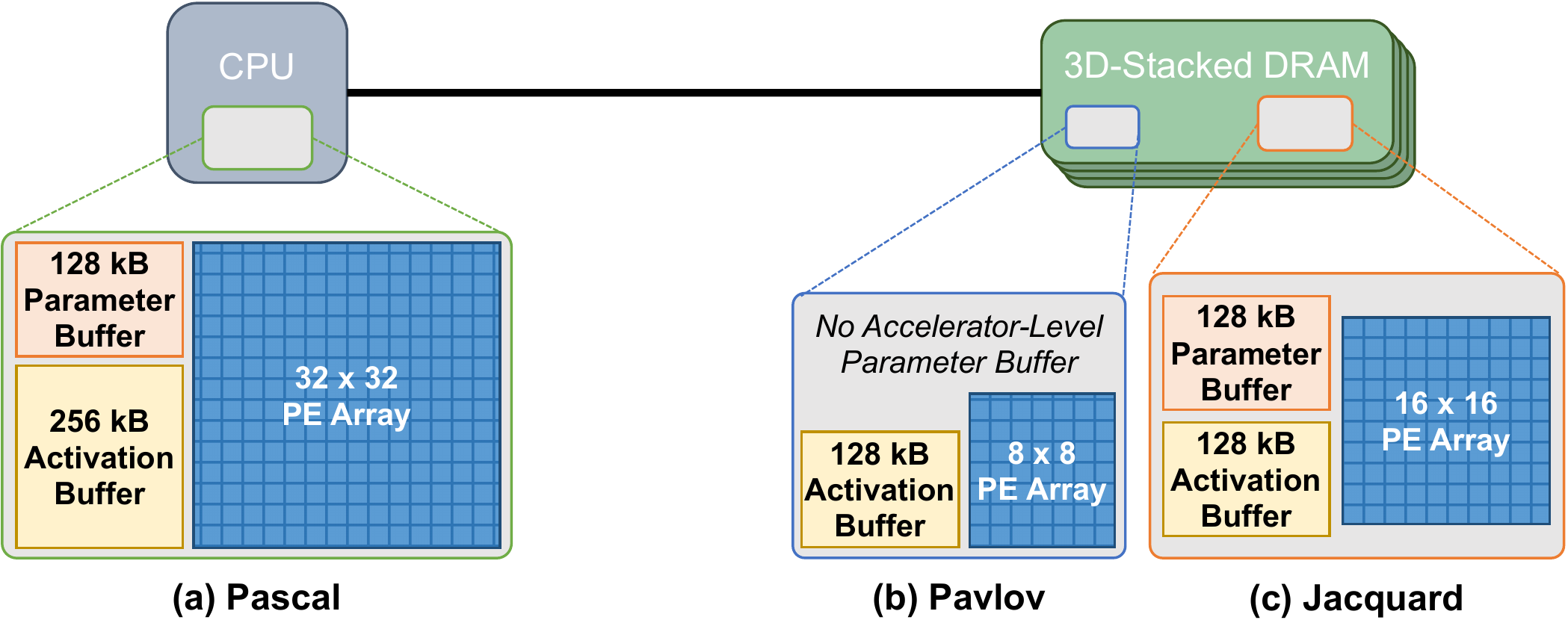}
    \caption{\exampleDesign accelerator design.}
    \label{fig:accel_all}
    \vspace{-10pt}
\end{figure}

\gf{Figure~\ref{fig:accel_all} shows the design of the three \titleShort accelerators, called  \accelA, \accelB, and \accelC, \gf{ which we collectively \gfi{refer} \gfcri{to} as \exampleDesign.} We tailor the hardware components (i.e., PE array size, on-chip parameter and activation buffers size, accelerator placement) and dataflow for each accelerator based on the properties of the layer families \gfcri{it} \gfcri{targets}. We briefly describe the hardware components of each accelerator next. We refer the reader to \gfcri{the full Mensa} paper~\cite{boroumand2021google} for a detailed description of each accelerator.}

\label{R1/6A}\Copy{R16A1}{\textbf{\accelA: Compute-Centric Accelerator Design.}
\accelA caters to layers in Families 1 and 2, which are compute-centric. We design \accelA as shown in Figure~\ref{fig:accel_all}a. First, the PE array in \accelA needs to have enough PEs to accommodate the high MAC intensity exhibited by layers in Families 1 and 2. Thus, we profile the performance of Family 1/2 layers on different PE sizes and empirically choose a 32$\times$32 PE array, which lets \accelA achieve a \SI{2}{\tera\flop\per\second} peak throughput. 
Second, we reduce the size of the activation buffer from \SI{2}{\mega\byte} in the Edge TPU to \SI{256}{\kilo\byte} \gfcri{(i.e., by 8$\times$)} in \accelA , because \accelA's dataflow exploits temporal reduction for the output activations using the internal PE registers, and no longer needs to store the large footprint of the output activations in the activation buffer. We reduce the size of the parameter buffer from \SI{4}{\mega\byte} in the Edge TPU to \SI{128}{\kilo\byte} in \accelA, because layers in Families 1 and 2 have small parameter footprints. Given the low off-chip memory bandwidth requirements of \accelA, we keep the accelerator on the CPU die. }

\Copy{R16A2}{\gfrev{\hl{\mbox{\accelA}'s dataflow reduces memory traffic by enabling two types of reuse. First, the dataflow uses temporal reduction~\gfcrii{\cite{meastro}} for each output activation element, by having a single PE accumulate the entire sum of the elements across multiple cycles in the PE's register file. Second, the dataflow uses spatial multicasting~\gfcrii{\cite{meastro}} for each parameter, by ensuring that all PEs \gfcrii{work} on the same \gfcri{image} channel\footnote{\gfcri{A layer in a CNN can sometimes break down an input into multiple filtered channels (e.g., breaking an image down into red, green, and blue colors).}} in the same cycle.}}}

\label{R1/6B}\Copy{R16B1}{\textbf{\accelB: \gfcri{Data-Centric Accelerator Design for (Mainly) LSTMs.}}
\accelB caters to layers in Family~3, which are data-centric \emph{and} predominantly consist of LSTM layers. We design \accelB as shown in Figure~\ref{fig:accel_all}b. 
\gf{First, b}ecause Family~3 layers have low MAC intensity, we empirically choose a small \gfcri{ 8$\times$8} PE array for \accelB. This allows \accelB to achieve \SI{128}{\giga\flop\per\second} peak throughput. 
\gf{Second, }we place \accelB \emph{inside memory} to accommodate the significant off-chip memory bandwidth requirements of Family~3 layers.
Given that parameters and activations from Family~3 layers exhibit different characteristics, for parameters, we use only one level of memory hierarchy (\SI{512}{\byte} of private registers per PE), eliminat\gfi{ing} the parameter buffer, and stream\gfi{ing} parameters directly from DRAM. The per-PE registers provide enough space to cache the temporally-multicasted parameters, and there are no other reuse opportunities that the parameter buffer could exploit.  Thanks to the small activation footprint of Family~3 layers, we use a \SI{128}{\kilo\byte} buffer for activations \gfcri{(16$\times$ reduction over \sg{the} Edge TPU's activation buffer)}.}

\Copy{R16B2}{\gfrev{\hl{\mbox{\accelB}'s dataflow reduces memory traffic by enabling two types of reuse. First, the dataflow temporally reuses a weight by allowing the PEs to store the weight and partial sums in one of their private registers. Second, the dataflow uses spatial multicasting for each input activation, as the same activation is multiplied across all columns for a given row \gfcri{of a parameter matrix}.}}}

\label{R1/6C}\Copy{R16C1}{\textbf{\gfcri{\accelC: Data-Centric Accelerator Design for (Mainly) Non-LSTM NNs.}}~\accelC caters to layers in Families 4 and 5, which primarily consist of non-LSTM data-centric layers. We design \accelC as shown in Figure~\ref{fig:accel_all}c.
\gf{First, w}hile layers in Families 4 and 5 have low MAC intensity, they perform more MAC operations on average than Family~3 layers. \gf{Thus, we empirically select a 16x16 PE array for \accelC.} This allows \accelC to achieve a peak throughput of \SI{512}{\giga\flop\per\second}.  
\gf{Second, }similar to \accelB, we place \accelC inside the logic layer of 3D-stacked memory. Doing so enables high memory bandwidth for the large parameter footprints of Family~4 layers. Given the small activation footprints, we use a small \SI{128}{\kilo\byte} buffer for them (a 16\gfi{$\times$} reduction compared to the Google Edge TPU). Thanks to the temporal parameter reuse that \accelC's dataflow enables, we reduce the parameter buffer to \SI{128}{\kilo\byte} (a 32\gfi{$\times$} reduction compared to the Google Edge TPU).}

\Copy{R16C2}{\gfrev{\hl{\mbox{\accelC}'s dataflow reduces memory traffic by enabling two types of reuse. First, the dataflow temporally reuses parameters. As in \mbox{\accelB}, a parameter is stored in a PE register and reused over multiple cycles to reduce the number of times the parameter is fetched from memory. By increasing the reuse of the parameter, the dataflow hides the off-chip memory access latency by overlapping it with PE computation. Second, the dataflow uses spatial multicasting for each input activation. To enable spatial multicasting for Families 4 and 5, the dataflow enables all PEs to collectively compute a single output activation in two steps. In the first step, each PE computes a partial sum. In the second step, the on-chip interconnect is used to gather partial sums from all PEs and produce the final output activation.}}
}

\label{R2/4A}\Copy{R2/4P1A}{\gfrev{\hl{\textbf{Layer-to-Accelerator Mapping.} We design a software runtime scheduler to identify which accelerator each layer in an NN model should run on. When an NN model runs on Mensa, the scheduler maps each NN layer to different accelerators. The scheduler uses the NN model (including a directed acyclic graph representing communication across model layers) and the configuration information in the accelerator hardware driver to determine this mapping.}}}

\Copy{R2/4P1B}{\gfrev{\hl{\textbf{Execution and Communication.} Once the layer-to-accelerator mapping is complete, Mensa begins model execution. During execution, destination layer $i$ placed in accelerator $j$ needs to read (1) any unbuffered parameters (i.e., weights) from DRAM; and (2) input activations (i.e., input data) produced by layer $i-1$, when layer $i$ is run on a different accelerator than $j$. In order to simplify communication between accelerators, Mensa accelerators transfer activations to another accelerator through DRAM, avoiding the need to keep on-chip data coherent across accelerators (or, when some of the Mensa accelerators are placed near memory, to keep on-chip and near-data accelerators coherent~\mbox{\gfcri{\cite{boroumand2021google, boroumand2018google,boroumand2019conda,amiraliphd}}}).}}} 

\subsection{Evaluation \gfrev{\hl{Methodology}}}

\label{R2/3B}\Copy{E3P2A}{We evaluate \gfrev{\hl{end-to-end}} inference energy \gf{(using an energy model based on prior works~\cite{boroumand2018google, gao2017tetris})}, hardware utilization\gf{,} and throughput \gf{(using an in-house simulator)}}\footnote{\Copy{E3P2B}{\gfrev{\label{E3B}\hl{We implement Mensa using Google's Edge TPU in-house simulator as a base, which faithfully models all major components of the Google Edge TPU, including the PE array, memory system, on-chip network, and dataflow. In our in-house simulator, the logic layer of 3D-stacked memory accelerators has access to the \SI{256}{\giga\byte\per\second} internal bandwidth of High Bandwidth Memory (HBM)~\cite{HBM}, which is 8$\times$ the external memory bandwidth \gfcri{of} accelerators that sit outside of memory.}}}\label{ft:E3B}}\Copy{E3P2C}{
of three configurations:
(1)~\emph{Baseline}, the Google Edge TPU~\gfcri{\cite{edge-tpu}};
(2)~\emph{Base+HB}, a hypothetical version of \base with 8$\times$ the memory bandwidth (\SI{256}{\giga\byte\per\second}), similar to \gfi{monolithic} 3D-\gfcri{stack}\gfi{-}based \gfcri{PNM} inference accelerators~\cite{gao2017tetris}; and
(3)~\emph{\exampleDesign} with all three accelerators (\accelA, \accelB, \accelC).
}

\subsection{Results}

\begin{figure}[ht]
       \vspace{-8pt}
    \centering
        \includegraphics[width=\linewidth]{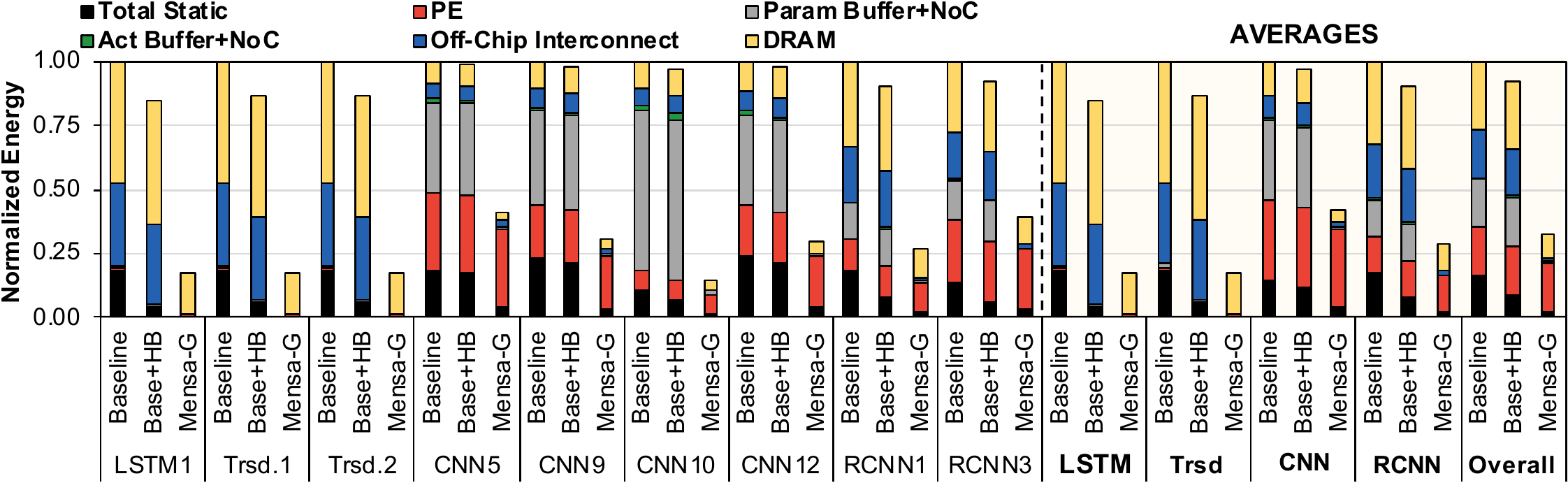}%
    \caption{Inference energy across different models normalized to Baseline.} 
       \vspace{-8pt}

    \label{fig:eval:energy-breakdown}
\end{figure}

\textbf{Energy Analysis.}
Figure~\ref{fig:eval:energy-breakdown} shows the total inference energy consumed by the three systems we evaluate across different NN models. \gf{We report results for representative LSTM, Transducers, CNNs, and RCNNs to improve the figure's clarity. We report average energy values for all 24 Google \edgemodels.} We make two observations. 

First, providing high memory bandwidth to \base (\basehb) results in only a small reduction in energy (7.5\% on average). This is because \basehb still incurs a high energy cost due to
(1)~on-chip buffers that are overprovisioned for many layers, and
(2)~off-chip traffic to DRAM. 
Second, \mensag significantly reduces energy across all models. The reduction primarily comes from three sources.
(1)~\mensag lowers the energy spent on on-chip and off-chip parameter traffic by 15.3$\times$ over \base, by scheduling each layer on the accelerator with the most appropriate \gfcri{design and} dataflow for that layer. LSTMs and Transducers benefit the most, as their energy in 
\basehb
is dominated by off-chip parameter traffic, which \accelB and \accelC drastically reduce by being placed inside memory.
(2)~\mensag reduces the dynamic energy of the on-chip buffer and \gfcri{on-chip} network (NoC) by 49.8$\times$ 
over \basehb 
by avoiding overprovisioning and catering to specialized dataflows. This is most beneficial for CNN and RCNN models.
(3)~\mensag reduces static energy by 3.6$\times$ 
over \basehb 
thanks to using significantly smaller PE arrays that avoid underutilization, significantly smaller on-chip buffers, and dataflows that reduce inference latency.

\begin{figure}[ht]
    \vspace{-8pt}
    \centering
        \centering
        \includegraphics[width=\linewidth]{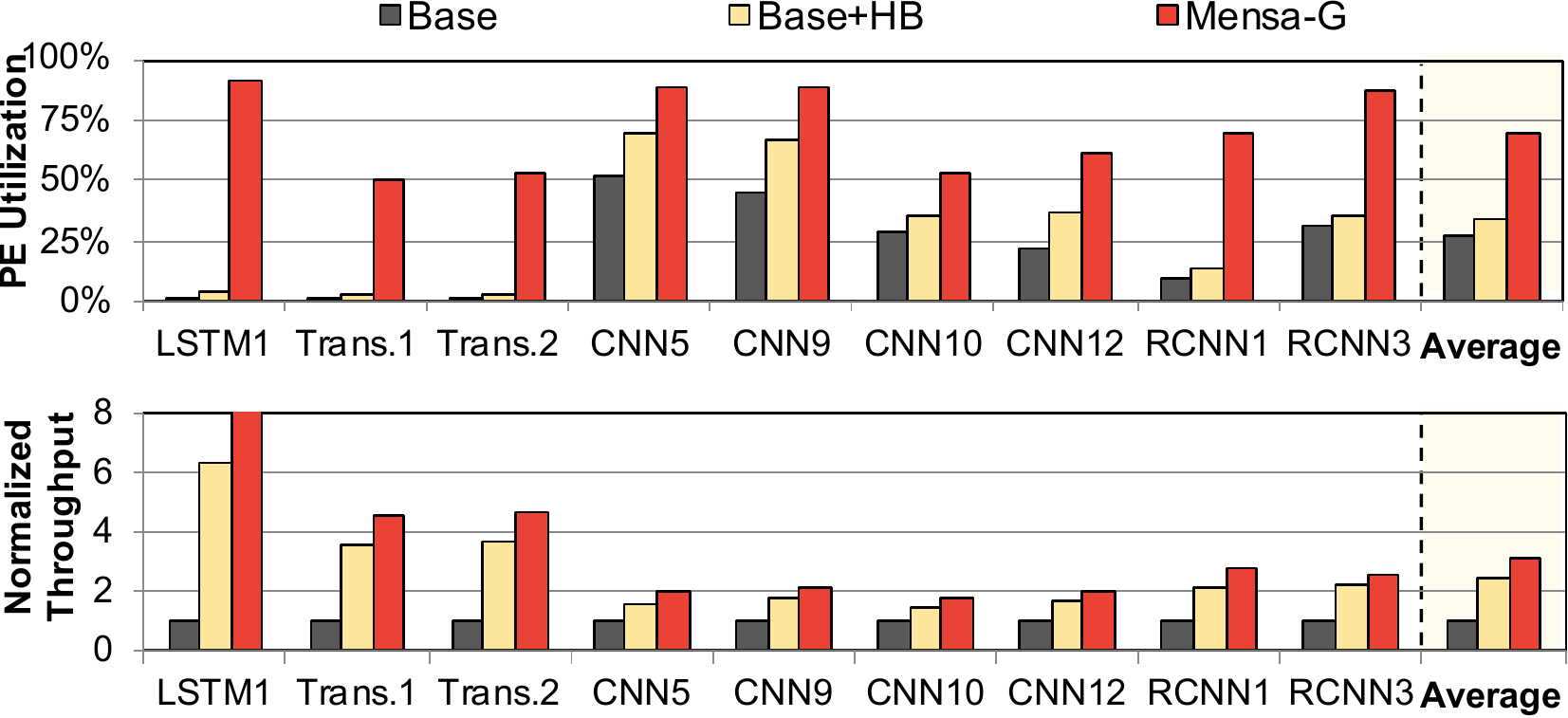}%
    \caption{PE utilization (top) and \base-normalized throughput (bottom).}
    \label{fig:eval:utilization}
\vspace{-8pt}

\end{figure}

\textbf{Utilization and Throughput Analyses.}
Figure~\ref{fig:eval:utilization} shows the raw PE utilization (top) and \base-normalized throughput (bottom) for our three configurations. \mensag's utilization is calculated by computing the average utilization across its three accelerators (\accelA, \accelB, and \accelC). We make two observations.  

First, \base suffers from low PE array utilization (on average 27.3\%). The higher memory bandwidth in \basehb increases the average PE utilization to 34.0\%, and improves throughput by 2.5$\times$. Overall, \basehb still has very low utilization, as many layers (those from Families 3, 4, and 5) do \emph{not} need the large number of PEs in the accelerator. 
Second, \mensag significantly increases both average utilization (by 2.5$\times$/2.0$\times$) and throughput (by 3.1$\times$/1.3$\times$) over \base/\basehb. The large utilization improvements are a result of
(1)~properly-provisioned PE arrays for each layer,
(2)~customized dataflows that exploit reuse and opportunities for parallelization, and
(3)~the movement of large-footprint layer computation into 3D-stacked memory (which eliminates off-chip traffic for their DRAM requests). 

\section{NN Inference on Processing-using-Memory}
\label{sec:simdram}

\gf{A common approach for \gfcri{PUM} architectures is to make use of bulk bitwise computation. Ambit~\gfcrii{\ambit}, a DRAM\gfi{-based} \gfcri{PUM} \gfi{architecture}, was the first work to propose exploiting DRAM's analog operation to perform bulk bitwise AND, OR, and NOT logic operations. Ambit leverages the fact that by simultaneously accessing three DRAM rows, the DRAM row sensing circuitry pulls its output voltage level to the majority voltage level among the three accessed rows, performing a majority-of-three (MAJ) operation. AND and OR logic operations can be performed by setting one of the three simultaneously accessed DRAM rows to either `1' (for an OR) or `0' (for a\gfi{n} AND). \gfcri{To implement NOT, Ambit uses and feeds into a DRAM array the complemented value of a cell that already exists after sensing in the sense amplifier.} \mech~\gfcri{\cite{hajinazarsimdram}} builds on top of Ambit to implement complex operations efficiently in DRAM using an end-to-end framework. Such \gfcri{a} framework efficiently implements arbitrary in-DRAM operations by 
\li~leveraging MAJ\gfi{/NOT}-based computation (instead of Ambit's \gfcri{focus on} AND/OR\gfi{/NOT}-based computation) and \lii~by employing a bit-serial execution model.} 

\label{R2/2A}\Copy{R2/2A}{\gf{SIMDRAM consists of three key steps to enable \gfcri{any} desired operation in DRAM: 
(1)~building an efficient MAJ/NOT-based representation of the desired operation, (2)~mapping the \gfcri{operation's} input and output operands to DRAM rows and to the required DRAM commands that produce the desired operation, and 
(3)~executing the operation. 
These three steps ensure efficient computation of a wide range of arbitrary and complex \gfcri{operations} in DRAM. The first two steps give users the flexibility to efficiently implement and compute any desired operation in DRAM. The third step controls the execution flow of the in-DRAM computation, transparently from the user.}}

\Copy{R2/2B}{\gfrev{\hl{For a desired computation, the first step derives its optimized MAJ/NOT-based implementation from its AND/OR/NOT-based implementation. The second step translates the MAJ/NOT-based implementation into DRAM row activations. This step includes (1) mapping the operands to the designated rows in DRAM, and (2) defining the sequence of DRAM row activations that are required to perform the computation. SIMDRAM chooses the operand-to-row mapping and the sequence of DRAM row activations to minimize the number of DRAM row activations required for a specific operation. To this end, we present a new \gfcri{row} allocation algorithm inspired by the linear scan register allocation algorithm~\gfcrii{\cite{poletto1999}}. However, unlike register allocation algorithms, our \gfcri{row} allocation algorithm considers two extra constraints that are specific to processing-using-DRAM: (1) performing MAJ in DRAM has destructive behavior; and (2) the number of DRAM rows that are designated to perform bitwise operations is limited. The third step programs the memory controller to issue the sequence of DRAM row activations to the appropriate rows in DRAM to perform the computation \gfcri{specified by} the operation. To this end, SIMDRAM uses a control unit in the memory controller that \gfcri{user-transparently} executes the sequence of DRAM row activations for each specific operation.}}}

\gf{We use SIMDRAM to efficiently support a wide range of operations of different types. We implement a set of 16 \mech operations of five different types for $n$-bit data elements: 
(1)~$N$-input logic operations (OR-/AND-/XOR-reduction across $N$ inputs); 
(2)~relational operations (equality/inequality check, greater-/less-than check, greater-than-or-equal-to check, and maximum/minimum element in a set); 
(3)~arithmetic operations (addition, subtraction, multiplication, division, and absolute value); 
(4)~predication (if-then-else); and 
(5)~other complex operations (bitcount, and ReLU~\gfcrii{\cite{goodfellow2016deep}}). 
We support four element sizes that correspond to data type sizes in popular programming languages (8-/16-/32-/64-bit integers). }

\subsection{\gfrev{\hl{System Integration}}}
\label{R2/4B}
\vspace{-5pt}

\gfrev{A program in a SIMDRAM-enabled system can have a combination of CPU and SIMDRAM instructions, with possible data sharing between the two. This leads to several challenges in integrating SIMDRAM into a real system. We discuss two system integration challenges related to data mapping and internal memory communication.} \gfrev{\hl{The full \gfcri{SIMDRAM} paper~\mbox{\cite{hajinazarsimdram}} contains further discussion about other system integration challenges and how we address them, including 
\li~data layout and how SIMDRAM \gfcri{stores} the data required for in-DRAM computation in a vertical layout \gfcri{and transposes such data when it is needed in a horizontal layout}; 
\lii~ISA extensions for and programming interface of SIMDRAM; 
\liii~how SIMDRAM handles page faults, address translation, coherence, and interrupts; 
\liv~how SIMDRAM manages computation on large amounts of data; 
\lv~security implications of SIMDRAM; and 
\lvi~current limitations \gfcri{of} SIMDRAM.}}\label{ft:R2/4P2B}

\Copy{R2/4P2C}{\gfrev{\textbf{\hl{Data Mapping.}} \hl{SIMDRAM uses a special ISA instruction (\texttt{bbop\_trsp\_init}) to inform the operating system (OS) that a memory location is a SIMDRAM object. This allows the OS to perform virtual-to-physical memory mapping optimizations for the SIMDRAM object before the allocation starts (e.g., mapping the arguments of an operation to the same row/column in physical memory).}}} 

\Copy{R2/4P2D}{\gfrev{\hl{\textbf{Internal Communication.}} \hl{SIMDRAM does \emph{not} support communication across different in-DRAM SIMD lanes (i.e., there is no communication across DRAM bitlines). Communication across subarrays and banks can be performed using prior works (i.e., LISA~\mbox{\cite{chang2016low}} and RowClone~\mbox{\cite{seshadri2013rowclone}}, respectively).}}}

\subsection{Evaluation Methodology}

\label{pg:E3A}\Copy{E3A}{\gf{We map three different binary neural networks (BNNs)~\gfcrii{\cite{rastegari2016xnor,lin2017towards,xiang2017binary,qin2020bipointnet,chen2021bnn}} to our \gfcri{PUM} substrate since SIMDRAM operates in a bit-serial model and only \gfrev{\hl{supports integer}} operations. BNNs are CNNs \gfrev{\hl{whose}} activations and weights are represented with 1-bit values, \gfcri{enabling} convolutions by executing bit-serial bitcount, shift, and addition operations~\cite{he2020sparse}. We use the XNOR-NET~\gfcrii{\cite{rastegari2016xnor}} implementations of VGG-13, VGG-16, and LeNET provided by \cite{he2020sparse}, to evaluate the functionality of SIMDRAM. We modify these implementations to make use of SIMDRAM's bitcount, addition, shift, and XNOR operations. We evaluate all three networks for inference using two different datasets:
\gfi{CIFAR-10~\gfcrii{\cite{krizhevsky2010convolutional}} (for VGG-13 and VGG-16), and MNIST~\gfcrii{\cite{deng2012mnist}} (for LeNet-5).}} \gfrev{\hl{We evaluate BNN inference time in SIMDRAM by isolating the main kernel that SIMDRAM executes (i.e., bitwise convolution) and evaluating its performance. We estimate the end-to-end speedup SIMDRAM provides for each BNN by applying Amdahl's law~\gfcrii{\cite{amdahl1967validity}}. Thus, SIMDRAM's end-to-end speedup is given by:  $((1 - conv\_time) + \frac{conv\_time}{SIMDRAM\_speedup})^{-1}$; 
where $SIMDRAM\_speedup$ is the speedup SIMDRAM provides for the main kernel compared to the CPU execution, and $conv\_time$ is the percentage of the total execution time the main kernel represents when executing BNN inference on the baseline CPU.}}}

\Copy{E3B}{We implement SIMDRAM using the gem5 simulator~\gfcrii{\cite{gem5}} and compare it to a real multicore CPU (Intel Skylake~\gfcrii{\cite{intelskylake}}
), a real high-end GPU (NVIDIA Titan V~\gfcrii{\cite{TitanV}}
), and a state-of-the-art processing-using-DRAM mechanism (Ambit~\gfcrii{\ambit}). \gfrev{\hl{We validate our Ambit and SIMDRAM gem5 implementations rigorously with the results reported in the original Ambit paper.}} We evaluate three different configurations of SIMDRAM where 1 (\emph{SIMDRAM:1}), 4 (\emph{SIMDRAM:4}), and 16 (\emph{SIMDRAM:16}) \gf{DRAM} banks out of all the 16 banks in one channel have SIMDRAM computation capability. \gfrev{\hl{\emph{SIMDRAM:1}'s throughput for bitcount, addition, shift, and XNOR operations used during \gfcrii{BNN} inference are 24.3 GOPs/s, 20.1 GOPs/s, 1337.5 GOPs/s, and 51.4 GOPs/s, respectively\gfcri{, and this throughput scales linearly with the number of DRAM banks}.}}}

\subsection{Results}

\begin{figure}[ht]
  \vspace{-5pt}
   \centering
   \includegraphics[width=\linewidth]{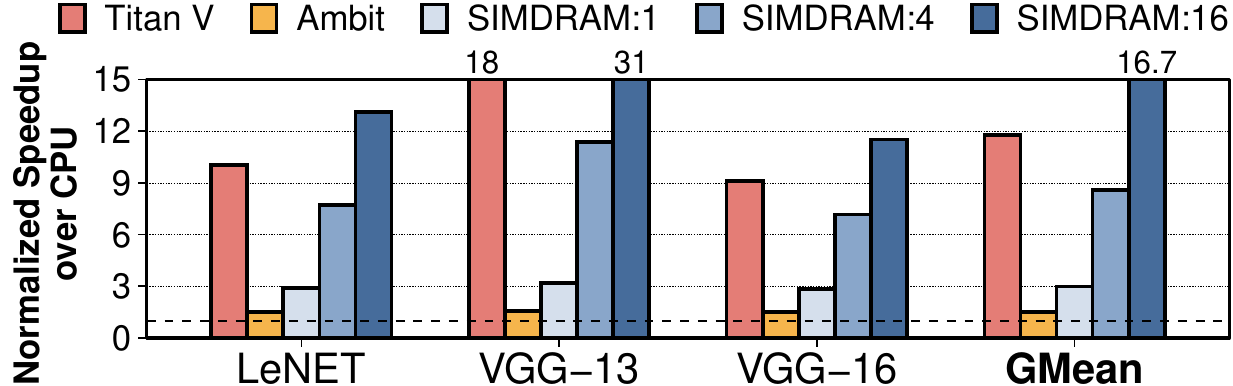}%
   \caption{Normalized speedup of 3 BNNs.}
   \label{fig_speedup_real_world}
   \vspace{-5pt}
\end{figure}

Figure~\ref{fig_speedup_real_world} shows the performance of \sg{\mech, the GPU, and Ambit} for each BNN, normalized to that of the multicore CPU. We make two observations. 

First, \emph{SIMDRAM:16} greatly outperforms the CPU and GPU baselines, providing 16.7$\times$ and 1.4$\times$ the performance of the CPU and GPU, respectively, on average across all three NNs. SIMDRAM has a maximum performance of 31$\times$ and 1.7$\times$ that of the CPU and GPU, respectively (for VGG-13 in both cases). Similarly, \emph{SIMDRAM:1} provides 1.9$\times$ the performance of Ambit (which also uses a single bank for in-DRAM computation), on average across all three NNs, with a maximum of 2$\times$ the performance of Ambit for VGG-13. 
Second, even with a single DRAM bank, SIMDRAM \emph{always} outperforms the CPU baseline, providing 3$\times$ the performance of the CPU on average across all NNs. 
We conclude that SIMDRAM is an effective and efficient substrate to accelerate \gfi{NN inference}.

\section{\gf{Key Takeaways} and Future Opportunities}
\label{sec:discussion}

This article investigates the benefits of state-of-the-art PIM architectures in mitigating memory bottlenecks during NN inference. We identify the key opportunities and challenges for three PIM solutions \gf{that can be employed in edge or cloud environments}. \gf{We summarize the key takeaways from our analysis as follows.}

\gfrev{\hl{First,}} \gf{we evaluate the performance of the GEMV primitive, which is \gfcri{a} key \gfcri{operation for} NN inference, on a commercial general-purpose server-scale \gfcri{PNM} architecture (UPMEM) using 32-bit floating-point and integer data types. Our experiments allow us to gain insights on the benefits a \emph{real} PIM architecture can provide for NN inference. We observe that while GEMV using \sg{floating point} takes one order of magnitude more time than using integers, the kernel execution time scales linearly with the number of UPMEM processing cores employed. When considering the 32-bit integer GEMV implementation, the UPMEM system provides similar performance to a high-end GPU, even outperforming the GPU when the GPU requires memory oversubscription. There are two key benefits of the UPMEM system compared to the other two PIM systems. (1)~It is relatively \emph{easy to scale} the UPMEM system to higher memory capacities and \sg{computational} power by simply adding more UPMEM \gfcri{modules} to \sg{a system}. In contrast, scaling the 3D-\sg{stack-}based \gfcri{PNM} architecture requires non-trivial engineering since 3D memory cubes need to be connected using specialized \gfcri{interconnection network}. (2)~It is relatively \emph{easy to program} the UPMEM system since UPMEM provides a complete programming toolkit and a C-based programming model for their system. In contrast, for the \gfcri{PUM} architecture, the programmer needs to manually map a workload to the underlying substrate, which is an arduous task.}

\gf{\gfrev{\hl{Second}}, Mensa, a heterogeneous architecture for Google \edgemodels that employs 3D-based \gfcri{PNM},  greatly benefits many NN models, improving performance (throughput and utilization) and energy efficiency\gfcri{, while at the same time reducing the area footprint of the monolithic Google Edge TPU}.
The key benefit of Mensa's 3D-based \gfcri{PNM} accelerator \gfcri{compared} to the 2D-based PIM designs we evaluate (SIMDRAM and UPMEM) is its \emph{flexibility}. 
Since the logic and memory components of a 3D-based architecture are separately manufactured, the design of \gfi{a} near-memory accelerator has no constraints other than the area and power \gfcri{constraints of} the logic layer. This means that, in practice, the 3D-based architecture can accelerate \gfcri{many NN types relatively flexibly}. In contrast, 2D \gfcri{PNM} and \gfcri{PUM} have natural limitations that make \gfcri{them infeasible or unsuitable}  to support floating-point computation, which is standard for most NN operations. The key drawback of \gfcri{PNM} solutions that rely on 3D-stacked memories is the \emph{high cost} and \emph{limited memory capacity} associated with 3D memories. This can be a limiting factor in edge environments (since edge devices are often cost-sensitive) and future NN models (since NN model sizes are increasing significantly).}

\gf{\gfrev{\hl{Third}}, SIMDRAM, a 2D-based \gfcri{PUM} architecture, provides significant performance improvements for binary neural networks compared to high-end CPUs and GPUs. The key benefit of employing a \gfcri{PUM} architecture for NN inference is the \emph{massive parallelism} available inside DRAM. SIMDRAM's throughput for a 32-bit addition using a single DRAM bank is 2.3$\times$ that of a CPU~\gfi{\cite{hajinazarsimdram}} and scales linearly with the number of DRAM banks (up to 36.8$\times$ the CPU throughput using all 16 DRAM banks in \gfi{a} DRAM chip~\gfi{\cite{hajinazarsimdram}}) \gfcri{with relatively low area cost (i.e., only 0.2\% area overhead in a high-end CPU)}. \mech is suitable for edge-based and cloud-based environments since it leverages the operating principles of standard DRAM DDRx memories. The key drawback of SIMDRAM is its \emph{limited applicability} for NN models due to the lack of 
\li~support for floating-point operations and 
\lii~tools (e.g., compilers, programming models) to map NN execution to the underlying hardware substrate.} 

We believe that there are many opportunities to improve the design of PIM architectures (both 2D and 3D) targeting NN inference, including better manufacturing processes that can \gfcri{seamlessly} integrate logic and memory units, cheaper and efficient interconnection networks that can be used to scale up the capacity of 3D-based  DRAM systems, and programming support to \gfcri{ease} the deployment of \gfcri{PUM} approaches. Also, there is a need for programming models and frameworks that can \emph{unify} the benefits of the different PIM architectures into a single heterogeneous system. Such solutions need to map, schedule, and control the execution of different NNs \gfcri{onto} the most appropriate PIM architecture, \gfcri{thereby} enabling a

\begin{NewBox}{s O{!htbp}}
\vspace{2pt}
\subsection{\textbf{Sidebar: Related Works on Accelerating NN Inference with Processing-in-DRAM}}
\vspace{2pt}

\textbf{2D \gfcri{PNM}.} \gfcri{Beyond UPMEM, other memory manufacturers have announced \gfcri{PNM} prototypes using 2D DRAM.} 
Samsung's AxDIMM~[\textcolor{blue}{1}] integrates reconfigurable logic to DDR4 \gfcri{modules} to accelerate recommendation systems. SK Hynix\gfcri{'s} GDDR6-based PIM~[\textcolor{blue}{2}] adds MAC units near DRAM banks to accelerate NN inference. Both solutions are not yet available \gfcri{on} the market. 

\textbf{3D \gfcri{PNM}.} There is extensive literature on accelerating NN inference using 3D-\gfcri{stack}-based \gfcri{PNM}~[\textcolor{blue}{3}]. We highlight Samsung's FIMDRAM~[\textcolor{blue}{4}], an industry proposal (not yet available \gfcri{on} the market) that adds MAC units near DRAM banks in an HBM2 device. A common drawback of prior 3D \gfcri{PNM} NN accelerators is that they use a monolithic accelerator design, which cannot efficiently accommodate diverse types of NNs. Mensa avoids such issues by tailoring NN accelerators for the needs of different NN models. 

\textbf{\gfcri{PUM}.} \gfcri{DRISA}~[\textcolor{blue}{5}] executes bulk operations in DRAM by adding logic elements inside DRAM arrays,  providing \gfcri{speedups} for binary neural networks. However, adding \gfcri{such} logic to the DRAM array incurs \gfcri{large} overheads. DrAcc~[\textcolor{blue}{6}] implements in-DRAM addition operations using AND/OR/NOT logic primitives, targeting ternary neural networks. SIMDRAM supports a wider range of operations (compared to DrAcc) at lower area overhead (compared to \gfcri{DRISA}). 

\vspace{5pt}
{\fontfamily{phv}\selectfont
\large{$\blacksquare$} REFERENCES
\footnotesize{
\begin{enumerate}[label={\arabic*.},leftmargin=0.5cm]

\item L. Ke \textit{et al.}, ``Near-Memory Processing in Action: Accelerating Personalized Recommendation with AxDIMM,`' \textit{IEEE Micro}, 2021.

\item S. Lee \textit{et al.}, ``A 1ynm 1.25V 8Gb, 16Gb/s/pin GDDR6-Based Accelerator-in-Memory Supporting 1 TFLOPS MAC Operation and Various Activation Functions for Deep-Learning Applications,`' in \textit{ISSCC}, 2022. 

\item A. S. Cordeiro \textit{et al.}, ``Efficient Machine Learning Execution with Near-Data Processing,`' \textit{MICPRO}, 2022.

\item Y-C. Kwon \textit{et al.}, ``A 20nm 6GB Function-in-Memory DRAM, Based on HBM2 with a 1.2 TFLOPS Programmable Computing Unit Using Bank-Level Parallelism, for Machine Learning Applications.`' in \textit{ISSCC}, 2021.

\item S. Li \textit{et al.}, ``\gfcri{DRISA}: A DRAM-Based Reconfigurable In-Situ Accelerator,`' in \textit{MICRO}, 2017.

\item Q. Deng \textit{et al.}, ``DrAcc: A DRAM Based Accelerator for Accurate CNN Inference,`' in \textit{DAC}, 2018.

\end{enumerate}
}
}
\end{NewBox}

\noindent workload to leverage the benefits of PIM while avoiding particular drawbacks related to a given PIM \gfcri{architecture}. We hope that our work brings \gfcri{attention} to such challenges and can inspire future PIM system designs.


\section{Conclusion}
\label{sec:conclusion}

We analyze, discuss, and contrast the benefits and drawbacks that different \gfii{DRAM-based} PIM architectures provide for NN performance and energy efficiency. To do so, we analyze three state-of-the-art PIM architectures, which broadly cover the PIM design space. Our comprehensive analysis provides several key takeaways. We conclude that while different PIM architectures provide greater performance and energy benefits than compute-centric solutions, different PIM architectures impose different limitations for NN workloads. We hope that our analysis can highlight the opportunities and drawbacks of state-of-the-art PIM solutions and inspire future designs.

\section{Acknowledgments}
We thank the anonymous reviewers of IEEE Micro for their valuable comments and feedback. We thank the SAFARI Research Group members for their valuable feedback and the stimulating intellectual environment they provide. We acknowledge support from the SAFARI Research Group's industrial partners, especially ASML, Facebook, Google, Huawei, Intel, Microsoft, and VMware. This work was supported in part by the Semiconductor Research Corporation (SRC) and the ETH Future Computing Laboratory.

\balance 
\setstretch{0.92}
\bibliographystyle{IEEEtran}
\bibliography{references}

\begin{thebibliography}{100}
\providecommand{\url}[1]{#1}
\csname url@samestyle\endcsname
\providecommand{\newblock}{\relax}
\providecommand{\bibinfo}[2]{#2}
\providecommand{\BIBentrySTDinterwordspacing}{\spaceskip=0pt\relax}
\providecommand{\BIBentryALTinterwordstretchfactor}{4}
\providecommand{\BIBentryALTinterwordspacing}{\spaceskip=\fontdimen2\font plus
\BIBentryALTinterwordstretchfactor\fontdimen3\font minus
  \fontdimen4\font\relax}
\providecommand{\BIBforeignlanguage}[2]{{%
\expandafter\ifx\csname l@#1\endcsname\relax
\typeout{** WARNING: IEEEtran.bst: No hyphenation pattern has been}%
\typeout{** loaded for the language `#1'. Using the pattern for}%
\typeout{** the default language instead.}%
\else
\language=\csname l@#1\endcsname
\fi
#2}}
\providecommand{\BIBdecl}{\relax}
\BIBdecl

\bibitem{rosenblatt1958perceptron}
F.~Rosenblatt, ``{The Perceptron: A Probabilistic Model for Information Storage
  and Organization in the Brain},'' \emph{Psychological Review}, 1958.

\bibitem{ivakhnenko1965}
A.~G. Ivakhnenko and V.~G. Lapa, \emph{{Cybernetic Predicting Devices}}.\hskip
  1em plus 0.5em minus 0.4em\relax CCM Information Corporation, 1965.

\bibitem{yegnanarayana2009artificial}
B.~Yegnanarayana, \emph{{Artificial Neural Networks}}.\hskip 1em plus 0.5em
  minus 0.4em\relax PHI Learning Pvt. Ltd., 2009.

\bibitem{oliveira2016computer}
T.~P. Oliveira, J.~S. Barbar, and A.~S. Soares, ``{Computer Network Traffic
  Prediction: A Comparison between Traditional and Deep Learning Neural
  Networks},'' \emph{IJBDI}, 2016.

\bibitem{al2011artificial}
Q.~K. Al-Shayea, ``{Artificial Neural Networks in Medical Diagnosis},''
  \emph{IJCSI}, 2011.

\bibitem{sze2017efficient}
V.~Sze, Y.-H. Chen \emph{et~al.}, ``{Efficient Processing of Deep Neural
  Networks: A Tutorial and Survey},'' \emph{Proceedings of the IEEE}, 2017.

\bibitem{fukushima.biologicalcybernetics1980}
K.~Fukushima, ``{Neocognitron: A Self-Organizing Neural Network Model for a
  Mechanism of Pattern Recognition Unaffected by Shift in Position},''
  \emph{Biological Cybernetics}, 1980.

\bibitem{lecun.cognitiva1985}
Y.~LeCun, ``{Une Proc\'{e}dure d'Apprentissage pour R\'{e}seau \`{a} Seuil
  Asym\'{e}trique},'' in \emph{Cognitiva}, 1985.

\bibitem{rumelhart.nature1986}
D.~Rumelhart, G.~E. Hinton, and R.~J. Williams, ``{Learning Representations by
  Back-Propagating Errors},'' \emph{Nature}, 1986.

\bibitem{lecun.nature2015}
Y.~LeCun, Y.~Bengio, and G.~Hinton, ``{Deep Learning},'' \emph{Nature}, 2015.

\bibitem{simonyan2015very}
K.~Simonyan and A.~Zisserman, ``{Very Deep Convolutional Networks for
  Large-Scale Image Recognition},'' in \emph{{ICLR}}, 2015.

\bibitem{gu2018recent}
J.~Gu, Z.~Wang \emph{et~al.}, ``{Recent Advances in Convolutional Neural
  Networks},'' \emph{Pattern Recognition}, 2018.

\bibitem{lecun1989handwritten}
Y.~LeCun, B.~Boser \emph{et~al.}, ``{Handwritten Digit Recognition with a
  Back-Propagation Network},'' in \emph{NeurIPS}, 1989.

\bibitem{lecun1998gradient}
Y.~LeCun, L.~Bottou \emph{et~al.}, ``{Gradient-Based Learning Applied to
  Document Recognition},'' \emph{Proc. IEEE}, 1998.

\bibitem{russakovsky2015imagenet}
O.~Russakovsky, J.~Deng \emph{et~al.}, ``{ImageNet Large Scale Visual
  Recognition Challenge},'' \emph{IJCV}, 2015.

\bibitem{zeiler2014visualizing}
M.~D. Zeiler and R.~Fergus, ``{Visualizing and Understanding Convolutional
  Networks},'' in \emph{ECCV}, 2014.

\bibitem{szegedy2015going}
C.~Szegedy, W.~Liu \emph{et~al.}, ``{Going Deeper with Convolutions},'' in
  \emph{CVPR}, 2015.

\bibitem{he.cvpr2016}
K.~He, X.~Zhang \emph{et~al.}, ``{Deep Residual Learning for Image
  Recognition},'' in \emph{CVPR}, 2016.

\bibitem{hochreiter.neco1997}
S.~Hochreiter and J.~Schmidhuber, ``{Long Short-Term Memory},'' \emph{NECO},
  1997.

\bibitem{gers.icann1999}
F.~Gers, J.~Schmidhuber, and F.~Cummins, ``{Learning to Forget: Continual
  Prediction with LSTM},'' in \emph{ICANN}, 1999.

\bibitem{greff.tnnls2017}
K.~Greff, R.~K. Srivastava \emph{et~al.}, ``{LSTM: A Search Space Odyssey},''
  \emph{TNNLS}, 2017.

\bibitem{lstm-google}
H.~Sak, A.~W. Senior, and F.~Beaufays, ``{Long Short-Term Memory Based
  Recurrent Neural Network Architectures for Large Vocabulary Speech
  Recognition},'' arXiv:1402.1128 [cs.NE], 2014.

\bibitem{graves2013generating}
A.~Graves, ``{Generating Sequences with Recurrent Neural Networks},''
  arXiv:1308.0850 [cs.NE], 2013.

\bibitem{google-translation}
Y.~Wu, M.~Schuster \emph{et~al.}, ``{Google's Neural Machine Translation
  System: Bridging the Gap Between Human and Machine Translation},''
  arXiv:1609.08144 [cs.CL], 2016.

\bibitem{cho2014learning}
K.~Cho, B.~Van~Merri{\"e}nboer \emph{et~al.}, ``{Learning Phrase
  Representations Using RNN Encoder-Decoder for Statistical Machine
  Translation},'' arXiv:1406.1078 [cs.CL], 2014.

\bibitem{lrcn}
J.~Donahue, L.~A. Hendricks \emph{et~al.}, ``{Long-Term Recurrent Convolutional
  Networks for Visual Recognition and Description},'' in \emph{CVPR}, 2015.

\bibitem{karpathy2015deep}
A.~Karpathy and L.~Fei-Fei, ``{Deep Visual-Semantic Alignments for Generating
  Image Descriptions},'' in \emph{CVPR}, 2015.

\bibitem{ranzato2014video}
M.~Ranzato, A.~Szlam \emph{et~al.}, ``{Video (Language) Modeling: A Baseline
  for Generative Models of Natural Videos},'' arXiv:1412.6604 [cs.LG], 2014.

\bibitem{srivastava2015unsupervised}
N.~Srivastava, E.~Mansimov, and R.~Salakhudinov, ``{Unsupervised Learning of
  Video Representations Using LSTMs},'' in \emph{ICML}, 2015.

\bibitem{sutskever2014sequence}
I.~Sutskever, O.~Vinyals, and Q.~V. Le, ``{Sequence to Sequence Learning with
  Neural Networks},'' in \emph{NeurIPS}, 2014.

\bibitem{xu2015show}
K.~Xu, J.~Ba \emph{et~al.}, ``{Show, Attend and Tell: Neural Image Caption
  Generation with Visual Attention},'' in \emph{ICML}, 2015.

\bibitem{xingjian2015convolutional}
S.~Xingjian, Z.~Chen \emph{et~al.}, ``{Convolutional LSTM Network: A Machine
  Learning Approach for Precipitation Nowcasting},'' in \emph{NeurIPS}, 2015.

\bibitem{gru}
J.~Chung, C.~Gulcehre \emph{et~al.}, ``{Empirical Evaluation of Gated Recurrent
  Neural Networks on Sequence Modeling},'' in \emph{NeurIPS}, 2014.

\bibitem{kanai2017preventing}
S.~Kanai, Y.~Fujiwara, and S.~Iwamura, ``{Preventing Gradient Explosions in
  Gated Recurrent Units},'' in \emph{NeurIPS}, 2017.

\bibitem{cho2014properties}
K.~Cho, B.~Van~Merri{\"e}nboer \emph{et~al.}, ``{On the Properties of Neural
  Machine Translation: Encoder--Decoder Approaches},'' arXiv:1409.1259 [cs.CL],
  2014.

\bibitem{graves.icmlworkshop2012}
A.~Graves, ``{Sequence Transduction with Recurrent Neural Networks},'' in
  \emph{ICML Representation Workshop}, 2012, arXiv:1211.3711 [cs.NE].

\bibitem{he.icassp2019}
Y.~He, T.~N. Sainath \emph{et~al.}, ``{Streaming End-to-End Speech Recognition
  for Mobile Devices},'' in \emph{ICASSP}, 2019.

\bibitem{liang.cvpr2015}
M.~Liang and X.~Hu, ``{Recurrent Convolutional Neural Network for Object
  Recognition},'' in \emph{CVPR}, 2015.

\bibitem{pinheiro.icml2014}
P.~Pinheiro and R.~Collobert, ``{Recurrent Convolutional Neural Networks for
  Scene Labeling},'' in \emph{ICML}, 2014.

\bibitem{rcnn-google}
O.~{Vinyals}, A.~{Toshev} \emph{et~al.}, ``{Show and Tell: A Neural Image
  Caption Generator},'' in \emph{CVPR}, 2015.

\bibitem{rastegari2016xnor}
M.~Rastegari, V.~Ordonez \emph{et~al.}, ``{XNOR-Net: ImageNet Classification
  Using Binary Convolutional Neural Networks},'' in \emph{ECCV}, 2016.

\bibitem{vanhoucke2011improving}
V.~Vanhoucke, A.~Senior, and M.~Z. Mao, ``{Improving the Speed of Neural
  Networks on CPUs},''
  \url{http://audentia-gestion.fr/Recherche-Research-Google/37631.pdf}, 2011.

\bibitem{jain2018architectural}
A.~Jain, M.~A. Laurenzano \emph{et~al.}, ``{Architectural Support for
  Convolutional Neural Networks on Modern CPUs},'' in \emph{PACT}, 2018.

\bibitem{adolf2016fathom}
R.~Adolf, S.~Rama \emph{et~al.}, ``{Fathom: Reference Workloads for Modern Deep
  Learning Methods},'' in \emph{IISWC}, 2016.

\bibitem{boroumand2021google}
A.~Boroumand, S.~Ghose \emph{et~al.}, ``{Google Neural Network Models for Edge
  Devices: Analyzing and Mitigating Machine Learning Inference Bottlenecks},''
  in \emph{PACT}, 2021.

\bibitem{tpu}
N.~P. Jouppi, C.~Young \emph{et~al.}, ``{In-Datacenter Performance Analysis of
  a Tensor Processing Unit},'' in \emph{ISCA}, 2017.

\bibitem{reddi2020mlperf}
V.~J. Reddi, C.~Cheng \emph{et~al.}, ``{MLPerf Inference Benchmark},'' in
  \emph{ISCA}, 2020.

\bibitem{wang2020neural}
S.~Wang, A.~Pathania, and T.~Mitra, ``{Neural Network Inference on Mobile
  SoCs},'' \emph{IEEE Design \& Test}, 2020.

\bibitem{wang2020systematic}
Y.~Wang, G.-Y. Wei, and D.~Brooks, ``{A Systematic Methodology for Analysis of
  Deep Learning Hardware and Software Platforms},'' in \emph{MLSys}, 2020.

\bibitem{wang2019benchmarking}
Y.~E. Wang, G.-Y. Wei, and D.~Brooks, ``{Benchmarking TPU, GPU, and CPU
  Platforms for Deep Learning},'' arXiv:1907.10701 [cs.LG], 2019.

\bibitem{gupta2019architectural}
U.~Gupta, X.~Wang \emph{et~al.}, ``{The Architectural Implications of
  Facebook's DNN-Based Personalized Recommendation},'' in \emph{HPCA}, 2020.

\bibitem{edge-tpu}
{Google LLC}, ``{Edge TPU},'' \url{https://cloud.google.com/edge-tpu/}.

\bibitem{chen2018tvm}
T.~Chen, T.~Moreau \emph{et~al.}, ``{$\{$TVM$\}$: An Automated
  $\{$End-to-End$\}$ Optimizing Compiler for Deep Learning},'' in \emph{OSDI},
  2018.

\bibitem{dnpu}
D.~{Shin}, J.~{Lee} \emph{et~al.}, ``{DNPU: An 8.1TOPS/W Reconfigurable
  CNN--RNN Processor for General-Purpose Deep Neural Networks},'' in
  \emph{ISSCC}, 2017.

\bibitem{eyeriss}
Y.-H. Chen, T.~Krishna \emph{et~al.}, ``{Eyeriss: An Energy-Efficient
  Reconfigurable Accelerator for Deep Convolutional Neural Networks},''
  \emph{JSSC}, 2017.

\bibitem{eyerissv2}
Y.-H. {Chen}, T.~{Yang} \emph{et~al.}, ``{Eyeriss v2: A Flexible Accelerator
  for Emerging Deep Neural Networks on Mobile Devices},'' \emph{JETCAS}, 2019.

\bibitem{scnn}
A.~{Parashar}, M.~{Rhu} \emph{et~al.}, ``{SCNN: An Accelerator for
  Compressed-Sparse Convolutional Neural Networks},'' in \emph{ISCA}, 2017.

\bibitem{han2016eie}
S.~Han, X.~Liu \emph{et~al.}, ``{EIE: Efficient Inference Engine on Compressed
  Deep Neural Network},'' in \emph{ISCA}, 2016.

\bibitem{kim2018deeptrain}
D.~Kim, T.~Na \emph{et~al.}, ``{DeepTrain: A Programmable Embedded Platform for
  Training Deep Neural Networks},'' \emph{TCAD}, 2018.

\bibitem{tangram}
M.~Gao, X.~Yang \emph{et~al.}, ``{TANGRAM: Optimized Coarse-Grained Dataflow
  for Scalable NN Accelerators},'' in \emph{ASPLOS}, 2019.

\bibitem{han2017ese}
S.~Han, J.~Kang \emph{et~al.}, ``{ESE: Efficient Speech Recognition Engine with
  Sparse LSTM on FPGA},'' in \emph{FPGA}, 2017.

\bibitem{lrcn-fpga}
X.~{Zhang}, X.~{Liu} \emph{et~al.}, ``{High-Performance Video Content
  Recognition with Long-Term Recurrent Convolutional Network for FPGA},'' in
  \emph{FPL}, 2017.

\bibitem{gao2017tetris}
M.~Gao, J.~Pu \emph{et~al.}, ``{TETRIS: Scalable and Efficient Neural Network
  Acceleration with 3D Memory},'' in \emph{ASPLOS}, 2017.

\bibitem{boroumand2018google}
A.~Boroumand, S.~Ghose \emph{et~al.}, ``{Google Workloads for Consumer Devices:
  Mitigating Data Movement Bottlenecks},'' in \emph{ASPLOS}, 2018.

\bibitem{imani2019floatpim}
M.~Imani, S.~Gupta \emph{et~al.}, ``{FloatPIM: In-Memory Acceleration of Deep
  Neural Network Training with High Precision},'' in \emph{ISCA}, 2019.

\bibitem{koppula2019eden}
S.~Koppula, L.~Orosa \emph{et~al.}, ``{EDEN: Enabling Energy-Efficient,
  High-Performance Deep Neural Network Inference Using Approximate DRAM},'' in
  \emph{MICRO}, 2019.

\bibitem{liu2018processing}
J.~Liu, H.~Zhao \emph{et~al.}, ``{Processing-in-Memory for Energy-Efficient
  Neural Network Training: A Heterogeneous Approach},'' in \emph{MICRO}, 2018.

\bibitem{min2019neuralhmc}
C.~Min, J.~Mao \emph{et~al.}, ``{NeuralHMC: An Efficient HMC-Based Accelerator
  for Deep Neural Networks},'' in \emph{ASP-DAC}, 2019.

\bibitem{cho2020mcdram}
S.~Cho, H.~Choi \emph{et~al.}, ``{McDRAM v2: In-Dynamic Random Access Memory
  Systolic Array Accelerator to Address the Large Model Problem in Deep Neural
  Networks on the Edge},'' \emph{IEEE Access}, 2020.

\bibitem{Shafiee2016}
A.~Shafiee, A.~Nag \emph{et~al.}, ``{ISAAC: A Convolutional Neural Network
  Accelerator with In-Situ Analog Arithmetic in Crossbars},'' in \emph{ISCA},
  2016.

\bibitem{eckert2018neural}
C.~Eckert, X.~Wang \emph{et~al.}, ``{Neural Cache: Bit-Serial In-Cache
  Acceleration of Deep Neural Networks},'' in \emph{ISCA}, 2018.

\bibitem{Chi2016}
P.~Chi, S.~Li \emph{et~al.}, ``{PRIME: A Novel Processing-in-Memory
  Architecture for Neural Network Computation in ReRAM-Based Main Memory},'' in
  \emph{ISCA}, 2016.

\bibitem{peemen2013memory}
M.~Peemen, A.~A. Setio \emph{et~al.}, ``{Memory-Centric Accelerator Design for
  Convolutional Neural Networks},'' in \emph{ICCD}, 2013.

\bibitem{ghose.ibmjrd19}
S.~Ghose, A.~Boroumand \emph{et~al.}, ``{Processing-in-Memory: A
  Workload-Driven Perspective},'' \emph{IBM JRD}, 2019.

\bibitem{mutlu2020modern}
O.~Mutlu, S.~Ghose \emph{et~al.}, ``{A Modern Primer on Processing in
  Memory},'' in \emph{Emerging Computing: From Devices to Systems --- Looking
  Beyond Moore and Von Neumann}.\hskip 1em plus 0.5em minus 0.4em\relax
  Springer, 2021.

\bibitem{deoliveira2021IEEE}
G.~F. Oliveira, J.~Gómez-Luna \emph{et~al.}, ``{DAMOV: A New Methodology and
  Benchmark Suite for Evaluating Data Movement Bottlenecks},'' \emph{IEEE
  Access}, 2021.

\bibitem{pim-book}
S.~Ghose, K.~Hsieh \emph{et~al.}, ``{The Processing-in-Memory Paradigm:
  Mechanisms to Enable Adoption},'' in \emph{Beyond-CMOS Technologies for Next
  Generation Computer Design}, 2019.

\bibitem{mutlu2019processing}
O.~Mutlu, S.~Ghose \emph{et~al.}, ``{Processing Data Where It Makes Sense:
  Enabling In-Memory Computation},'' \emph{MicPro}, 2019.

\bibitem{mutlu2019enabling}
O.~Mutlu, S.~Ghose \emph{et~al.}, ``{Enabling Practical Processing in and Near
  Memory for Data-Intensive Computing},'' in \emph{DAC}, 2019.

\bibitem{mutlu2015research}
O.~Mutlu and L.~Subramanian, ``{Research Problems and Opportunities in Memory
  Systems},'' \emph{SUPERFRI}, 2014.

\bibitem{mutlu2013memory}
O.~Mutlu, ``{Memory Scaling: A Systems Architecture Perspective},'' in
  \emph{IMW}, 2013.

\bibitem{loh2013processing}
G.~H. Loh, N.~Jayasena \emph{et~al.}, ``{A Processing in Memory Taxonomy and a
  Case for Studying Fixed-Function PIM},'' in \emph{WoNDP}, 2013.

\bibitem{Near-Data}
R.~Balasubramonian, J.~Chang \emph{et~al.}, ``{Near-Data Processing: Insights
  from a MICRO-46 Workshop},'' \emph{IEEE Micro}, 2014.

\bibitem{stone1970logic}
H.~S. Stone, ``{A Logic-in-Memory Computer},'' \emph{IEEE Trans. Comput.},
  1970.

\bibitem{Miss_Mem_Wall_1996}
A.~Saulsbury, F.~Pong, and A.~Nowatzyk, ``{Missing the Memory Wall: The Case
  for Processor/Memory Integration},'' in \emph{ISCA}, 1996.

\bibitem{farmahini2015nda}
A.~Farmahini-Farahani, J.~H. Ahn \emph{et~al.}, ``{NDA: Near-DRAM Acceleration
  Architecture Leveraging Commodity DRAM Devices and Standard Memory
  Modules},'' in \emph{HPCA}, 2015.

\bibitem{babarinsa2015jafar}
O.~O. Babarinsa and S.~Idreos, ``{JAFAR: Near-Data Processing for Databases},''
  in \emph{SIGMOD}, 2015.

\bibitem{devaux2019true}
F.~Devaux, ``{The True Processing in Memory Accelerator},'' in \emph{Hot
  Chips}, 2019.

\bibitem{ghiasi2022genstore}
N.~M. Ghiasi, J.~Park \emph{et~al.}, ``{GenStore: A High-Performance and
  Energy-Efficient In-Storage Computing System for Genome Sequence Analysis},''
  in \emph{ASPLOS}, 2022.

\bibitem{gomez2021benchmarkingcut}
J.~G{\'o}mez-Luna, I.~El~Hajj \emph{et~al.}, ``{Benchmarking Memory-Centric
  Computing Systems: Analysis of Real Processing-in-Memory Hardware},'' in
  \emph{CUT}, 2021.

\bibitem{gomezluna2021benchmarking}
J.~G{\'o}mez-Luna, I.~E. Hajj \emph{et~al.}, ``{Benchmarking a New Paradigm: An
  Experimental Analysis of a Real Processing-in-Memory Architecture},''
  arXiv:2105.03814 [cs.AR], 2021.

\bibitem{gomez2022benchmarking}
J.~G{\'o}mez-Luna, I.~El~Hajj \emph{et~al.}, ``{Benchmarking a New Paradigm:
  Experimental Analysis and Characterization of a Real Processing-in-Memory
  System},'' \emph{IEEE Access}, 2022.

\bibitem{syncron}
C.~Giannoula, N.~Vijaykumar \emph{et~al.}, ``{SynCron: Efficient
  Synchronization Support for Near-Data-Processing Architectures},'' in
  \emph{HPCA}, 2021.

\bibitem{singh2020nero}
G.~Singh, D.~Diamantopoulos \emph{et~al.}, ``{NERO: A Near High-Bandwidth
  Memory Stencil Accelerator for Weather Prediction Modeling},'' in \emph{FPL},
  2020.

\bibitem{skhynixpim}
S.~Lee, K.~Kim \emph{et~al.}, ``{A 1ynm 1.25V 8Gb, 16Gb/s/pin GDDR6-based
  Accelerator-in-Memory Supporting 1TFLOPS MAC Operation and Various Activation
  Functions for Deep-Learning Applications},'' in \emph{ISSCC}, 2022.

\bibitem{ke2021near}
L.~Ke, X.~Zhang \emph{et~al.}, ``{Near-Memory Processing in Action:
  Accelerating Personalized Recommendation with AxDIMM},'' \emph{IEEE Micro},
  2021.

\bibitem{giannoula2022sparsep}
C.~Giannoula, I.~Fernandez \emph{et~al.}, ``{SparseP: Towards Efficient Sparse
  Matrix Vector Multiplication on Real Processing-in-Memory Architectures},''
  in \emph{SIGMETRICS}, 2022.

\bibitem{shin2018mcdram}
H.~Shin, D.~Kim \emph{et~al.}, ``{McDRAM: Low Latency and Energy-Efficient
  Matrix Computations in DRAM},'' \emph{IEEE TCADICS}, 2018.

\bibitem{denzler2021casper}
A.~Denzler, R.~Bera \emph{et~al.}, ``{Casper: Accelerating Stencil Computation
  using Near-Cache Processing},'' arXiv:2112.14216 [cs.AR], 2021.

\bibitem{asghari2016chameleon}
H.~Asghari-Moghaddam, Y.~H. Son \emph{et~al.}, ``{Chameleon: Versatile and
  Practical Near-DRAM Acceleration Architecture for Large Memory Systems},'' in
  \emph{MICRO}, 2016.

\bibitem{IRAM_Micro_1997}
D.~Patterson, T.~Anderson \emph{et~al.}, ``{A Case for Intelligent RAM},''
  \emph{IEEE Micro}, 1997.

\bibitem{C_RAM_1999}
D.~G. Elliott, M.~Stumm \emph{et~al.}, ``{Computational RAM: Implementing
  Processors in Memory},'' \emph{Design and Test of Computers}, vol.~16, Jan.
  1999.

\bibitem{CASES_MVX}
M.~A.~Z. Alves, P.~C. Santos \emph{et~al.}, ``Saving memory movements through
  vector processing in the dram,'' in \emph{Int. Conf. on Compilers,
  Architecture and Synthesis for Embedded Systems}, 2015.

\bibitem{Xi_2015}
S.~L. Xi, O.~Babarinsa \emph{et~al.}, ``Beyond the wall: Near-data processing
  for databases,'' in \emph{Int. Workshop on Data Management on New Hardware},
  2015.

\bibitem{sun2021abc}
W.~Sun, Z.~Li \emph{et~al.}, ``{ABC-DIMM: Alleviating the Bottleneck of
  Communication in DIMM-Based Near-Memory Processing with Inter-DIMM
  Broadcast},'' in \emph{ISCA}, 2021.

\bibitem{matam2019graphssd}
K.~K. Matam, G.~Koo \emph{et~al.}, ``{GraphSSD: Graph Semantics Aware SSD},''
  in \emph{ISCA}, 2019.

\bibitem{gokhale1995processing}
M.~Gokhale, B.~Holmes, and K.~Iobst, ``{Processing in Memory: The Terasys
  Massively Parallel PIM Array},'' \emph{Computer}, 1995.

\bibitem{hall1999mapping}
M.~Hall, P.~Kogge \emph{et~al.}, ``{Mapping Irregular Applications to DIVA, a
  PIM-Based Data-Intensive Architecture},'' in \emph{SC}, 1999.

\bibitem{MEMSYS_MVX}
M.~A.~Z. Alves, P.~C. Santos \emph{et~al.}, ``{Opportunities and Challenges of
  Performing Vector Operations Inside the DRAM},'' in \emph{MEMSYS}, 2015.

\bibitem{lockerman2020livia}
E.~Lockerman, A.~Feldmann \emph{et~al.}, ``{Livia: Data-Centric Computing
  Throughout the Memory Hierarchy},'' in \emph{ASPLOS}, 2020.

\bibitem{ahn2015scalable}
J.~Ahn, S.~Hong \emph{et~al.}, ``{A Scalable Processing-in-Memory Accelerator
  for Parallel Graph Processing},'' in \emph{ISCA}, 2015.

\bibitem{nai2017graphpim}
L.~Nai, R.~Hadidi \emph{et~al.}, ``{GraphPIM: Enabling Instruction-Level PIM
  Offloading in Graph Computing Frameworks},'' in \emph{HPCA}, 2017.

\bibitem{lazypim}
A.~Boroumand, S.~Ghose \emph{et~al.}, ``{LazyPIM: An Efficient Cache Coherence
  Mechanism for Processing-in-Memory},'' \emph{CAL}, 2017.

\bibitem{top-pim}
D.~Zhang, N.~Jayasena \emph{et~al.}, ``{TOP-PIM: Throughput-Oriented
  Programmable Processing in Memory},'' in \emph{HPDC}, 2014.

\bibitem{gao2016hrl}
M.~Gao and C.~Kozyrakis, ``{HRL: Efficient and Flexible Reconfigurable Logic
  for Near-Data Processing},'' in \emph{HPCA}, 2016.

\bibitem{kim2018grim}
J.~S. Kim, D.~S. Cali \emph{et~al.}, ``{GRIM-Filter: Fast Seed Location
  Filtering in DNA Read Mapping Using Processing-in-Memory Technologies},''
  \emph{BMC Genomics}, 2018.

\bibitem{drumond2017mondrian}
M.~Drumond, A.~Daglis \emph{et~al.}, ``{The Mondrian Data Engine},'' in
  \emph{ISCA}, 2017.

\bibitem{RVU}
P.~C. Santos, G.~F. Oliveira \emph{et~al.}, ``{Operand Size Reconfiguration for
  Big Data Processing in Memory},'' in \emph{DATE}, 2017.

\bibitem{NIM}
G.~F. Oliveira, P.~C. Santos \emph{et~al.}, ``{NIM: An HMC-Based Machine for
  Neuron Computation},'' in \emph{ARC}, 2017.

\bibitem{PEI}
J.~Ahn, S.~Yoo \emph{et~al.}, ``{PIM-Enabled Instructions: A Low-Overhead,
  Locality-Aware Processing-in-Memory Architecture},'' in \emph{ISCA}, 2015.

\bibitem{Kim2016}
D.~Kim, J.~Kung \emph{et~al.}, ``{Neurocube: A Programmable Digital
  Neuromorphic Architecture with High-Density 3D Memory},'' in \emph{ISCA},
  2016.

\bibitem{gu2016leveraging}
P.~Gu, S.~Li \emph{et~al.}, ``{Leveraging 3D Technologies for Hardware
  Security: Opportunities and Challenges},'' in \emph{GLSVLSI}, 2016.

\bibitem{boroumand2019conda}
A.~Boroumand, S.~Ghose \emph{et~al.}, ``{CoNDA: Efficient Cache Coherence
  Support for Near-Data Accelerators},'' in \emph{ISCA}, 2019.

\bibitem{hsieh2016transparent}
K.~Hsieh, E.~Ebrahimi \emph{et~al.}, ``{Transparent Offloading and Mapping
  (TOM) Enabling Programmer-Transparent Near-Data Processing in GPU Systems},''
  in \emph{ISCA}, 2016.

\bibitem{cali2020genasm}
D.~S. Cali, G.~S. Kalsi \emph{et~al.}, ``{GenASM: A High-Performance, Low-Power
  Approximate String Matching Acceleration Framework for Genome Sequence
  Analysis},'' in \emph{MICRO}, 2020.

\bibitem{NDC_ISPASS_2014}
S.~H. Pugsley, J.~Jestes \emph{et~al.}, ``{NDC: Analyzing the Impact of
  3D-Stacked Memory+Logic Devices on MapReduce Workloads},'' in \emph{ISPASS},
  2014.

\bibitem{pattnaik2016scheduling}
A.~Pattnaik, X.~Tang \emph{et~al.}, ``{Scheduling Techniques for GPU
  Architectures with Processing-in-Memory Capabilities},'' in \emph{PACT},
  2016.

\bibitem{akin2015data}
B.~Akin, F.~Franchetti, and J.~C. Hoe, ``{Data Reorganization in Memory Using
  3D-Stacked DRAM},'' in \emph{ISCA}, 2015.

\bibitem{hsieh2016accelerating}
K.~Hsieh, S.~Khan \emph{et~al.}, ``{Accelerating Pointer Chasing in 3D-Stacked
  Memory: Challenges, Mechanisms, Evaluation},'' in \emph{ICCD}, 2016.

\bibitem{lee2015bssync}
J.~H. Lee, J.~Sim, and H.~Kim, ``{BSSync: Processing Near Memory for Machine
  Learning Workloads with Bounded Staleness Consistency Models},'' in
  \emph{PACT}, 2015.

\bibitem{boroumand2021mitigating}
A.~Boroumand, S.~Ghose \emph{et~al.}, ``{Mitigating Edge Machine Learning
  Inference Bottlenecks: An Empirical Study on Accelerating Google Edge
  Models},'' arXiv:2103.00768 [cs.AR], 2021.

\bibitem{boroumand2022polynesia}
A.~Boroumand, S.~Ghose \emph{et~al.}, ``{Polynesia: Enabling High-Performance
  and Energy-Efficient Hybrid Transactional/Analytical Databases with
  Hardware/Software Co-Design},'' in \emph{ICDE}, 2022.

\bibitem{boroumand2021polynesia}
A.~Boroumand, S.~Ghose \emph{et~al.}, ``{Polynesia: Enabling Effective Hybrid
  Transactional/Analytical Databases with Specialized Hardware/Software
  Co-Design},'' arXiv:2103.00798 [cs.AR], 2021.

\bibitem{amiraliphd}
A.~Boroumand, ``{Practical Mechanisms for Reducing Processor-Memory Data
  Movement in Modern Workloads},'' Ph.D. dissertation, Carnegie Mellon
  University, 2020.

\bibitem{besta2021sisa}
M.~Besta, R.~Kanakagiri \emph{et~al.}, ``Sisa: Set-centric instruction set
  architecture for graph mining on processing-in-memory systems,'' in
  \emph{MICRO}, 2021.

\bibitem{fernandez2020natsa}
I.~Fernandez, R.~Quislant \emph{et~al.}, ``{NATSA: A Near-Data Processing
  Accelerator for Time Series Analysis},'' in \emph{ICCD}, 2020.

\bibitem{singh2019napel}
G.~Singh, G.~ \emph{et~al.}, ``{NAPEL: Near-Memory Computing Application
  Performance Prediction via Ensemble Learning},'' in \emph{DAC}, 2019.

\bibitem{kwon202125}
Y.-C. Kwon, S.~H. Lee \emph{et~al.}, ``{A 20nm 6GB Function-in-Memory DRAM,
  Based on HBM2 with a 1.2 TFLOPS Programmable Computing Unit using Bank-Level
  Parallelism, for Machine Learning Applications},'' in \emph{ISSCC}, 2021.

\bibitem{lee2021hardware}
S.~Lee, S.-h. Kang \emph{et~al.}, ``{Hardware Architecture and Software Stack
  for PIM Based on Commercial DRAM Technology: Industrial Product},'' in
  \emph{ISCA}, 2021.

\bibitem{niu2022184qps}
D.~Niu, S.~Li \emph{et~al.}, ``{184QPS/W 64Mb/$mm^2$ 3D Logic-to-DRAM Hybrid
  Bonding with Process-Near-Memory Engine for Recommendation System},'' in
  \emph{ISSCC}, 2022.

\bibitem{Sparse_MM_LiM}
Q.~Zhu, T.~Graf \emph{et~al.}, ``{Accelerating Sparse Matrix-Matrix
  Multiplication with 3D-Stacked Logic-in-Memory Hardware},'' in \emph{HPEC},
  2013.

\bibitem{azarkhish2016logic}
E.~Azarkhish, C.~Pfister \emph{et~al.}, ``{Logic-Base Interconnect Design for
  Near Memory Computing in the Smart Memory Cube},'' \emph{IEEE VLSI}, 2016.

\bibitem{azarkhish2018neurostream}
E.~Azarkhish, D.~Rossi \emph{et~al.}, ``{Neurostream: Scalable and Energy
  Efficient Deep Learning with Smart Memory Cubes},'' \emph{TPDS}, 2018.

\bibitem{guo20143d}
Q.~Guo, N.~Alachiotis \emph{et~al.}, ``{3D-Stacked Memory-Side Acceleration:
  Accelerator and System Design},'' in \emph{WoNDP}, 2014.

\bibitem{de2018design}
J.~P.~C. de~Lima, P.~C. Santos \emph{et~al.}, ``{Design Space Exploration for
  PIM Architectures in 3D-Stacked Memories},'' in \emph{CF}, 2018.

\bibitem{akin2014hamlet}
B.~Ak{\i}n, J.~C. Hoe, and F.~Franchetti, ``{HAMLeT: Hardware Accelerated
  Memory Layout Transform within 3D-Stacked DRAM},'' in \emph{HPEC}, 2014.

\bibitem{huang2020heterogeneous}
Y.~Huang, L.~Zheng \emph{et~al.}, ``{A Heterogeneous PIM Hardware-Software
  Co-Design for Energy-Efficient Graph Processing},'' in \emph{IPDPS}, 2020.

\bibitem{dai2018graphh}
G.~Dai, T.~Huang \emph{et~al.}, ``{GraphH: A Processing-in-Memory Architecture
  for Large-Scale Graph Processing},'' \emph{TCAD}, 2018.

\bibitem{tsai:micro:2018:ams}
P.-A. Tsai, C.~Chen, and D.~Sanchez, ``{Adaptive Scheduling for Systems with
  Asymmetric Memory Hierarchies},'' in \emph{MICRO}, 2018.

\bibitem{gu2020ipim}
P.~Gu, X.~Xie \emph{et~al.}, ``{iPIM: Programmable In-Memory Image Processing
  Accelerator using Near-Bank Architecture},'' in \emph{ISCA}, 2020.

\bibitem{DRAMA_CAL_2014}
A.~Farmahini-Farahani, J.~H. Ahn \emph{et~al.}, ``{DRAMA: An Architecture for
  Accelerated Processing Near Memory},'' \emph{Computer Architecture Letters},
  2014.

\bibitem{Asghari-Moghaddam_2016}
H.~Asghari-Moghaddam, A.~Farmahini-Farahani \emph{et~al.}, ``{Near-DRAM
  Acceleration with Single-ISA Heterogeneous Processing in Standard Memory
  Modules},'' \emph{IEEE Micro}, 2016.

\bibitem{huang2019active}
J.~Huang, R.~R. Puli \emph{et~al.}, ``{Active-Routing: Compute on the Way for
  Near-Data Processing},'' in \emph{HPCA}, 2019.

\bibitem{kersey2017lightweight}
C.~D. Kersey, H.~Kim, and S.~Yalamanchili, ``{Lightweight SIMT Core Designs for
  Intelligent 3D Stacked DRAM},'' in \emph{MEMSYS}, 2017.

\bibitem{li2019pims}
J.~Li, X.~Wang \emph{et~al.}, ``{PIMS: A Lightweight Processing-in-Memory
  Accelerator for Stencil Computations},'' in \emph{MEMSYS}, 2019.

\bibitem{kim2017grim}
J.~S. Kim, D.~Senol \emph{et~al.}, ``{GRIM-Filter: Fast Seed Filtering in Read
  Mapping using Emerging Memory Technologies},'' arXiv:1708.04329 [q-bio.GN],
  2017.

\bibitem{boroumand2017lazypim}
A.~Boroumand, S.~Ghose \emph{et~al.}, ``{LazyPIM: Efficient Support for Cache
  Coherence in Processing-in-Memory Architectures},'' arXiv:1706.03162 [cs.AR],
  2017.

\bibitem{zhuo2019graphq}
Y.~Zhuo, C.~Wang \emph{et~al.}, ``{GraphQ: Scalable PIM-Based Graph
  Processing},'' in \emph{MICRO}, 2019.

\bibitem{zhang2018graphp}
M.~Zhang, Y.~Zhuo \emph{et~al.}, ``{GraphP: Reducing Communication for
  PIM-Based Graph Processing with Efficient Data Partition},'' in \emph{HPCA},
  2018.

\bibitem{lim2017triple}
H.~Lim and G.~Park, ``{Triple Engine Processor (TEP): A Heterogeneous
  Near-Memory Processor for Diverse Kernel Operations},'' \emph{TACO}, 2017.

\bibitem{smc_sim}
E.~Azarkhish, D.~Rossi \emph{et~al.}, ``{A Case for Near Memory Computation
  Inside the Smart Memory Cube},'' in \emph{EMS}, 2016.

\bibitem{HIVE}
M.~A.~Z. Alves, M.~Diener \emph{et~al.}, ``{Large Vector Extensions Inside the
  HMC},'' in \emph{DATE}, 2016.

\bibitem{jang2019charon}
J.~Jang, J.~Heo \emph{et~al.}, ``{Charon: Specialized Near-Memory Processing
  Architecture for Clearing Dead Objects in Memory},'' in \emph{MICRO}, 2019.

\bibitem{IBM_ActiveCube}
R.~Nair, S.~F. Antao \emph{et~al.}, ``{Active Memory Cube: A
  Processing-in-Memory Architecture for Exascale Systems},'' \emph{IBM JRD},
  2015.

\bibitem{hadidi2017cairo}
R.~Hadidi, L.~Nai \emph{et~al.}, ``{CAIRO: A Compiler-Assisted Technique for
  Enabling Instruction-Level Offloading of Processing-in-Memory},''
  \emph{TACO}, 2017.

\bibitem{santos2018processing}
P.~C. Santos, G.~F. Oliveira \emph{et~al.}, ``{Processing in 3D Memories to
  Speed Up Operations on Complex Data Structures},'' in \emph{DATE}, 2018.

\bibitem{seshadri2017ambit}
V.~Seshadri, D.~Lee \emph{et~al.}, ``{Ambit: In-Memory Accelerator for Bulk
  Bitwise Operations Using Commodity DRAM Technology},'' in \emph{MICRO}, 2017.

\bibitem{seshadri2019dram}
V.~Seshadri and O.~Mutlu, ``{In-DRAM Bulk Bitwise Execution Engine},''
  arXiv:1905.09822 [cs.AR], 2019.

\bibitem{li2017drisa}
S.~Li, D.~Niu \emph{et~al.}, ``{DRISA: A DRAM-Based Reconfigurable In-Situ
  Accelerator},'' in \emph{MICRO}, 2017.

\bibitem{seshadri2013rowclone}
V.~Seshadri, Y.~Kim \emph{et~al.}, ``{RowClone: Fast and Energy-Efficient
  In-DRAM Bulk Data Copy and Initialization},'' in \emph{MICRO}, 2013.

\bibitem{seshadri2016processing}
V.~Seshadri and O.~Mutlu, ``{The Processing Using Memory Paradigm: In-DRAM Bulk
  Copy, Initialization, Bitwise AND and OR},'' arXiv:1610.09603 [cs.AR], 2016.

\bibitem{deng2018dracc}
Q.~Deng, L.~Jiang \emph{et~al.}, ``{DrAcc: A DRAM Based Accelerator for
  Accurate CNN Inference},'' in \emph{DAC}, 2018.

\bibitem{xin2020elp2im}
X.~Xin, Y.~Zhang, and J.~Yang, ``{ELP2IM: Efficient and Low Power Bitwise
  Operation Processing in DRAM},'' in \emph{HPCA}, 2020.

\bibitem{song2018graphr}
L.~Song, Y.~Zhuo \emph{et~al.}, ``{GraphR: Accelerating Graph Processing Using
  ReRAM},'' in \emph{HPCA}, 2018.

\bibitem{song2017pipelayer}
L.~Song, X.~Qian \emph{et~al.}, ``{PipeLayer: A Pipelined ReRAM-Based
  Accelerator for Deep Learning},'' in \emph{HPCA}, 2017.

\bibitem{gao2019computedram}
F.~Gao, G.~Tziantzioulis, and D.~Wentzlaff, ``{ComputeDRAM: In-Memory Compute
  Using Off-the-Shelf DRAMs},'' in \emph{MICRO}, 2019.

\bibitem{aga2017compute}
S.~Aga, S.~Jeloka \emph{et~al.}, ``{Compute Caches},'' in \emph{HPCA}, 2017.

\bibitem{dualitycache}
D.~Fujiki, S.~Mahlke, and R.~Das, ``{Duality Cache for Data Parallel
  Acceleration},'' in \emph{ISCA}, 2019.

\bibitem{seshadri2016buddy}
V.~Seshadri, D.~Lee \emph{et~al.}, ``{Buddy-RAM: Improving the Performance and
  Efficiency of Bulk Bitwise Operations Using DRAM},'' arXiv:1611.09988
  [cs.AR], 2016.

\bibitem{seshadri.bookchapter17}
V.~Seshadri and O.~Mutlu, ``{Simple Operations in Memory to Reduce Data
  Movement},'' in \emph{Advances in Computers, Volume 106}, 2017.

\bibitem{seshadri2018rowclone}
V.~Seshadri, Y.~Kim \emph{et~al.}, ``{RowClone: Accelerating Data Movement and
  Initialization Using DRAM},'' arXiv:1805.03502 [cs.AR], 2018.

\bibitem{seshadri2015fast}
V.~Seshadri, K.~Hsieh \emph{et~al.}, ``{Fast Bulk Bitwise AND and OR in
  DRAM},'' \emph{CAL}, 2015.

\bibitem{li2016pinatubo}
S.~Li, C.~Xu \emph{et~al.}, ``{Pinatubo: A Processing-in-Memory Architecture
  for Bulk Bitwise Operations in Emerging Non-Volatile Memories},'' in
  \emph{DAC}, 2016.

\bibitem{ferreira2021pluto}
J.~D. Ferreira, G.~Falcao \emph{et~al.}, ``{pLUTo: In-DRAM Lookup Tables to
  Enable Massively Parallel General-Purpose Computation},'' arXiv:2104.07699
  [cs.AR], 2021.

\bibitem{ferreira2022pluto}
J.~D. Ferreira, G.~Falcao \emph{et~al.}, ``{pLUTo: Enabling Massively Parallel
  Computation in DRAM via Lookup Tables},'' in \emph{MICRO}, 2022.

\bibitem{he2020sparse}
Z.~He, L.~Yang \emph{et~al.}, ``{Sparse BD-Net: A Multiplication-Less DNN with
  Sparse Binarized Depth-Wise Separable Convolution},'' \emph{JETC}, 2020.

\bibitem{flashcosmos}
J.~Park, R.~Azizi \emph{et~al.}, ``{Flash-Cosmos: In-Flash Bulk Bitwise
  Operations Using Inherent Computation Capability of NAND Flash Memory},'' in
  \emph{MICRO}, 2022.

\bibitem{truong2022adapting}
M.~S. Truong, L.~Shen \emph{et~al.}, ``{Adapting the RACER Architecture to
  Integrate Improved In-ReRAM Logic Primitives},'' \emph{JETCAS}, 2022.

\bibitem{truong2021racer}
M.~S. Truong, E.~Chen \emph{et~al.}, ``{RACER: Bit-Pipelined Processing using
  Resistive Memory},'' in \emph{MICRO}, 2021.

\bibitem{olgun2021quactrng}
A.~Olgun, M.~Patel \emph{et~al.}, ``{QUAC-TRNG: High-Throughput True Random
  Number Generation Using Quadruple Row Activation in Commodity DRAMs},'' in
  \emph{ISCA}, 2021.

\bibitem{kim2019d}
J.~S. Kim, M.~Patel \emph{et~al.}, ``{D-RaNGe: Using Commodity DRAM Devices to
  Generate True Random Numbers With Low Latency and High Throughput},'' in
  \emph{HPCA}, 2019.

\bibitem{kim2018dram}
J.~S. Kim, M.~Patel \emph{et~al.}, ``{The DRAM Latency PUF: Quickly Evaluating
  Physical Unclonable Functions by Exploiting the Latency-Reliability Tradeoff
  in Modern Commodity DRAM Devices},'' in \emph{HPCA}, 2018.

\bibitem{bostanci2022dr}
F.~N. Bostanc{\i}, A.~Olgun \emph{et~al.}, ``{DR-STRaNGe: End-to-End System
  Design for DRAM-Based True Random Number Generators},'' in \emph{HPCA}, 2022.

\bibitem{olgun2022pidram}
A.~Olgun, J.~G. Luna \emph{et~al.}, ``{PiDRAM: A Holistic End-to-End FPGA-Based
  Framework for Processing-in-DRAM},'' \emph{TACO}, 2022.

\bibitem{ali2019memory}
M.~F. Ali, A.~Jaiswal, and K.~Roy, ``{In-Memory Low-Cost Bit-Serial Addition
  Using Commodity DRAM Technology},'' in \emph{{TCAS-I}}, 2019.

\bibitem{angizi2019graphide}
S.~Angizi and D.~Fan, ``{GraphiDe: A Graph Processing Accelerator Leveraging
  In-DRAM-Computing},'' in \emph{GLSVLSI}, 2019.

\bibitem{li2018scope}
S.~Li, A.~O. Glova \emph{et~al.}, ``{SCOPE: A Stochastic Computing Engine for
  DRAM-Based In-Situ Accelerator},'' in \emph{MICRO}, 2018.

\bibitem{subramaniyan2017parallel}
A.~Subramaniyan and R.~Das, ``{Parallel Automata Processor},'' in \emph{ISCA},
  2017.

\bibitem{zha2020hyper}
Y.~Zha and J.~Li, ``{Hyper-AP: Enhancing Associative Processing Through A
  Full-Stack Optimization},'' in \emph{ISCA}, 2020.

\bibitem{fujiki2018memory}
D.~Fujiki, S.~Mahlke, and R.~Das, ``{In-Memory Data Parallel Processor},'' in
  \emph{ASPLOS}, 2018.

\bibitem{orosa2021codic}
L.~Orosa, Y.~Wang \emph{et~al.}, ``{CODIC: A Low-Cost Substrate for Enabling
  Custom In-DRAM Functionalities and Optimizations},'' in \emph{ISCA}, 2021.

\bibitem{sharad2013ultra}
M.~Sharad, D.~Fan, and K.~Roy, ``{Ultra Low Power Associative Computing with
  Spin Neurons and Resistive Crossbar Memory},'' in \emph{DAC}, 2013.

\bibitem{rezaei2020nom}
S.~H.~S. Rezaei, M.~Modarressi \emph{et~al.}, ``{NoM: Network-on-Memory for
  Inter-Bank Data Transfer in Highly-Banked Memories},'' \emph{CAL}, 2020.

\bibitem{upmem}
UPMEM, ``{UPMEM Website},'' \url{https://www.upmem.com}, 2020.

\bibitem{upmem2018}
UPMEM, ``{Introduction to UPMEM PIM. Processing-in-memory (PIM) on DRAM
  Accelerator (White Paper)},'' 2018.

\bibitem{hajinazarsimdram}
N.~Hajinazar, G.~F. Oliveira \emph{et~al.}, ``{SIMDRAM: A Framework for
  Bit-Serial SIMD Processing Using DRAM},'' in \emph{ASPLOS}, 2021.

\bibitem{williams2009roofline}
S.~Williams, A.~Waterman, and D.~Patterson, ``{Roofline: An Insightful Visual
  Performance Model for Multicore Architectures},'' \emph{CACM}, 2009.

\bibitem{energy-roofline-model}
J.~W. {Choi}, D.~{Bedard} \emph{et~al.}, ``{A Roofline Model of Energy},'' in
  \emph{IPDPS}, 2013.

\bibitem{lin2017towards}
X.~Lin, C.~Zhao, and W.~Pan, ``{Towards Accurate Binary Convolutional Neural
  Network},'' in \emph{NIPS}, 2017.

\bibitem{xiang2017binary}
X.~Xiang, Y.~Qian, and K.~Yu, ``{Binary Deep Neural Networks for Speech
  Recognition},'' in \emph{INTERSPEECH}, 2017.

\bibitem{qin2020bipointnet}
H.~Qin, Z.~Cai \emph{et~al.}, ``{BiPointNet: Binary Neural Network for Point
  Clouds},'' \emph{arXiv:2010.05501 [cs.CV]}, 2020.

\bibitem{chen2021bnn}
T.~Chen, Z.~Zhang \emph{et~al.}, ``{" BNN-BN=?": Training Binary Neural
  Networks Without Batch Normalization},'' in \emph{CVPR}, 2021.

\bibitem{a100}
{NVIDIA}, ``{NVIDIA A100 Tensor Core GPU Architecture. White Paper},''
  \url{https://images.nvidia.com/aem-dam/en-zz/Solutions/data-center/nvidia-ampere-architecture-whitepaper.pdf},
  2020.

\bibitem{choquette2020nvidia}
J.~Choquette and W.~Gandhi, ``{NVIDIA A100 GPU: Performance \& Innovation for
  GPU Computing},'' in \emph{Hot Chips}, 2020.

\bibitem{cublas}
NVIDIA, ``{CUDA Basic Linear Algebra Subroutine (cuBLAS) Library},''
  \url{https://docs.nvidia.com/cuda/cublas/index.html}, 2022.

\bibitem{hbm2}
{JEDEC Solid State Technology Assn.}, ``{JESD23-5D: High Bandwidth Memory (HBM)
  DRAM Standard},'' March 2021.

\bibitem{kim1996assessing}
Y.-B. Kim and T.~W. Chen, ``{Assessing Merged DRAM/Logic Technology},'' in
  \emph{ISCA}, 1996.

\bibitem{li2019framework}
C.~Li, R.~Ausavarungnirun \emph{et~al.}, ``{A Framework for Memory
  Oversubscription Management in Graphics Processing Units},'' in
  \emph{ASPLOS}, 2019.

\bibitem{ausavarungnirun2017mosaic}
R.~Ausavarungnirun, J.~Landgraf \emph{et~al.}, ``{Mosaic: A GPU Memory Manager
  with Application-Transparent Support for Multiple Page Sizes},'' in
  \emph{MICRO}, 2017.

\bibitem{ausavarungnirun2018mask}
R.~Ausavarungnirun, V.~Miller \emph{et~al.}, ``{MASK: Redesigning the GPU
  Memory Hierarchy to Support Multi-Application Concurrency},'' in
  \emph{ASPLOS}, 2018.

\bibitem{HBM}
D.~U. Lee, K.~W. Kim \emph{et~al.}, ``{A 1.2V 8Gb 8-Channel 128GB/s
  High-Bandwidth Memory (HBM) Stacked DRAM with Effective Microbump I/O Test
  Methods Using 29nm Process and TSV},'' in \emph{ISSCC}, 2014.

\bibitem{HMC2}
{Hybrid Memory Cube Consortium}, ``{Hybrid Memory Cube Specification Rev.
  2.0},'' \url{http://www.hybridmemorycube.org/}.

\bibitem{lee2016simultaneous}
D.~Lee, S.~Ghose \emph{et~al.}, ``{Simultaneous Multi-Layer Access: Improving
  3D-Stacked Memory Bandwidth at Low Cost},'' \emph{TACO}, 2016.

\bibitem{meastro}
H.~Kwon, P.~Chatarasi \emph{et~al.}, ``{Understanding Reuse, Performance, and
  Hardware Cost of DNN Dataflow: A Data-Centric Approach},'' in \emph{MICRO},
  2019.

\bibitem{poletto1999}
M.~Poletto and V.~Sarkar, ``{Linear Scan Register Allocation},'' \emph{TOPLAS},
  1999.

\bibitem{goodfellow2016deep}
I.~Goodfellow, Y.~Bengio, and A.~Courville, \emph{{Deep Learning}}.\hskip 1em
  plus 0.5em minus 0.4em\relax MIT Press, 2016.

\bibitem{chang2016low}
K.~K. Chang, P.~J. Nair \emph{et~al.}, ``{Low-Cost Inter-Linked Subarrays
  (LISA): Enabling Fast Inter-Subarray Data Movement in DRAM},'' in
  \emph{HPCA}, 2016.

\bibitem{krizhevsky2010convolutional}
A.~Krizhevsky, ``{Convolutional Deep Belief Networks on CIFAR-10},''
  \url{https://www.cs.toronto.edu/~kriz/conv-cifar10-aug2010.pdf}, 2010.

\bibitem{deng2012mnist}
L.~Deng, ``{The MNIST Database of Handwritten Digit Images for Machine Learning
  Research [Best of the Web]},'' \emph{IEEE Signal Processing Magazine}, 2012.

\bibitem{amdahl1967validity}
G.~M. Amdahl, ``{Validity of the Single Processor Approach to Achieving Large
  Scale Computing Capabilities},'' in \emph{AFIPS}, 1967.

\bibitem{gem5}
N.~Binkert, B.~Beckmann \emph{et~al.}, ``{The gem5 Simulator},'' \emph{Comput.
  Archit. News}, 2011.

\bibitem{intelskylake}
{Intel Corp.}, ``{6th Generation Intel Core Processor Family Datasheet},''
  \url{http://www.intel.com/content/www/us/en/processors/core/}.

\bibitem{TitanV}
{NVIDIA Corp.}, ``{NVIDIA Titan V},''
  \url{https://www.nvidia.com/en-us/titan/titan-v/}.

\end{thebibliography}
\setstretch{1.0}

\begin{IEEEbiography}{Geraldo F. Oliveira} is currently with ETH Zürich, Zürich, Switzerland. Contact him at geraldod@safari.ethz.ch.
\end{IEEEbiography}

\begin{IEEEbiography}{Juan Gómez-Luna} is currently with ETH Zürich, Zürich, Switzerland. He is a Member of IEEE. Contact him at juan.gomez@safari.ethz.ch.
\end{IEEEbiography}

\begin{IEEEbiography}{Saugata Ghose} is currently with the University of Illinois Urbana\gfcri{-}Champaign, Urbana, IL, USA. He is a Member of IEEE. Contact him at ghose@illinois.edu.
\end{IEEEbiography}

\begin{IEEEbiography}{Amirali Boroumand} is currently with Google, Mountain View, CA, USA. Contact him at amirali.boroumand@gmail.com.
\end{IEEEbiography}

\begin{IEEEbiography}{Onur Mutlu} is currently with ETH Zürich, Zürich, Switzerland and Carnegie Mellon University, Pittsburgh, PA, USA. He is a
Fellow of IEEE. Contact him at omutlu@gmail.com. 
\end{IEEEbiography}

\end{document}